\documentclass{aastex63}

\usepackage{amsmath}
\usepackage{xcolor}

\newcommand{{\HII}}{H\,{\sc ii} }
\newcommand{{\HI}}{H\,{\sc i} }

\newcommand{\um}{\,\(\mu\)m }
\newcommand{\kms}{\,km s\textsuperscript{-1}}
\newcommand{\wno}{\,cm\textsuperscript{-1}}
\newcommand{\amin}{ arcmin}
\newcommand{\flux}{ erg s\textsuperscript{-1} cm\textsuperscript{-2} sr\textsuperscript{-1} cm}

\newcommand\tilded{{\raise.17ex\hbox{$\scriptstyle\sim$}}}
\usepackage{graphicx}
\received{August 3, 2021}
\accepted{MNRAS \today}

\shorttitle{ Kinematic Structure of W33 Main's UCHII Regions}
\shortauthors{Beilis et al.}

\bibpunct[, ]{(}{)}{;}{a}{,}{,}

\setcitestyle{aysep={}} 

\begin{document}

\title{ Kinematics and Structure of Ionized Gas in the UCHII Regions of W33 Main}

\correspondingauthor{Dan Beilis}
\email{danbeilis@mail.tau.ac.il}

\author{Dan Beilis}
\affiliation{Raymond and Beverly Sackler School of Physics and Astronomy \\
	Tel Aviv University \\
	Tel Aviv, 69978, Israel}

\author{Sara Beck}
\affiliation{Raymond and Beverly Sackler School of Physics and Astronomy \\
Tel Aviv University \\
Tel Aviv, 69978, Israel}

\author{John Lacy}
\affiliation{Department of Astronomy \\
University of Texas \\
Austin, TX 78712, USA}

\begin{abstract}

High mass proto-stars create Ultra-Compact \HII regions (UCHII)  at the stage of evolution when most of the accretion is finished but the star is still heavily embedded in molecular material.  
The morphologies of UCHII regions reflect the interactions of stellar winds, stellar motions, and density structure in the molecular cloud; they are complex and it has been very difficult to interpret them.   We here present data obtained with TEXES on the NASA IRTF of the [NeII] emission line in the proto-cluster of young OB stars in W33 Main.  The data cube has a spatial resolution of $\sim1.4$ arcsec and true velocity resolution $\sim5$\kms; with $A_\lambda\sim0.02A$\textsubscript{V} it is relatively unaffected by extinction.   We have run 3D hydrodynamic and line profile simulations, using PLUTO and RADMC-3D, of the gas structures created by multiple windy stars moving relative to each other through the ambient cloud.  By iterative comparison of the data cube and the simulations, we arrive at models that reproduce the different morphology and kinematic structure of each UCHII region in W33 Main.  The results indicate that each sub-source probably holds multiple exciting stars, permitting an improved view of the stellar population, and finds the stellar trajectories,  which may determine the dynamical development of the proto-cluster.
\end{abstract}


\section{Introduction} \label{sec:intro}
Young Stellar Objects (YSOs) of $M_*\gtrsim8$M\textsubscript{\(\odot\)} evolve very fast and can become strong sources of heat, luminosity and ionization while still deeply embedded in the natal molecular cloud.   They cannot be seen at optical, UV or the shorter IR wavelengths because of the high extinction. They are most readily observed via the Ultra-Compact \HII~ Regions (UCHII) they excite.  UCHII regions are defined by \citet{wood1989morphologies}, to have diameters $\leq0.1$ pc, emission measures $\geq$$10^7$ pc cm\textsuperscript{-6} and electron densities $\geq$$10^4$ cm\textsuperscript{-3}.      Stars spend on order \tilded $10^5$ yr, much longer than the free expansion time, in this stage: thus presenting the 'lifetime problem' in UCHII regions.     This is a crucial stage in star formation as UCHII regions can have significant effects on the surrounding clouds. 

 Radio continuum observations of UCHII regions find complex morphologies, classified by \citet{wood1989morphologies}  as spherical, cometary, irregular, core-halo and shell (\citep*{kurtz1994ultracompact}; \citet{walsh1998studies}).    It is not usually possible to deduce how a given UCHII region has evolved and will develop from its appearance alone;  to disentangle all the possible mechanisms needs kinematic data as well.   Data on the gas kinematics in UCHII regions is however very limited. Optical tracers cannot be used because of the high extinction. Radio recombination lines of \HI are observable, but are limited in velocity resolution: thermal broadening of \HI in a $10^4$ K \HII~region is $\simeq20$\kms,  within a factor of 2 of the typical FWHM  observed.   The mid-infrared fine structure emission lines of common metals have proven to be useful kinematic tracers in obscured \HII regions.  They have the advantage over \HI recombination lines that they are broadened less by thermal motions, and they are much less affected by extinction than any optical emission lines.   \citep{Zhu_2008,Jaffe_2003} used the 12.8\um line of [NeII] to probe the internal gas motions in some UCHII regions and \citet{Beck_2015, Beck_2020}  used [NeII] 12.8\um and [SIV]10.5\um to study gas motions in embedded extragalactic super star clusters.  This paper presents high resolution observations of the UCHII regions in the Galactic star-forming region W33-Main in the form of a full  3-D cube of the [NeII] line with $\sim1.4$ arcsec~ spatial and  $\sim5$\kms~ velocity resolution.
    
Many mechanisms including star-cloud interactions, structures in the ambient cloud, stellar winds and stellar motions can affect the appearance and kinematics of the UCHII region. In this paper we focus on the effects of ionizing, windy stars moving through the cloud.  We use  PLUTO and RADMC-3D to produce a suite of  3-D hydrodynamic simulations and  simulated line profiles of moving stars,  which we interactively tweak to create models of each UCHII region in W33 that reproduce the observed spatial and kinematic structures in the data cube.   We review W33 and the observations in the next section, the  numerical simulation methods in Section 3, and the results in Section 4.  In Sections 5 we discuss the implication of our findings for the embedded stellar population and in Section 6 the potential value of this method.  These simulations show the effects of multiplicity in each source and the trajectories of the included stars and suggest the future evolution of the proto-cluster.

\section{Target and Observations} \label{sec:observations_data_reduction}

\subsection{W33 Main} \label{sec:w33_main}

W33 is a giant molecular cloud complex in the Galaxy’s Scutum spiral arm \citep{immer2013trigonometric} at a distance $\sim2.4$ kpc; 1 arcsec~ $\sim$ $0.12$ pc on the source.  In it lies a $\sim10 $ pc  star formation region including all evolutionary stages from inactive clumps to fully developed \HII regions \citep{immer2014diversity}.
  W33 Main is the central molecular clump in the W33 complex. It has molecular mass of \tilded 3965 M\textsubscript{\(\odot\)} \citep{immer2014diversity} and multiple signs of active massive star formation, including water and methanol masers \citep*{genzel1977h2o,haschick1990detection}. 
 The molecular gas in W33 has two velocity components at $+35$ and $+58$\kms~ \citep{Kohno_2018} and these two velocities are seen also in the H$134$\(\alpha\) recombination line measured over the full $10$\amin~extent of the clouds \citep*{Gardner_1975}. It holds a small cluster of embedded \HII regions detected in the infrared at 3.4-33\um by \citet{dyck1977infrared}   and in the radio by  \citet{haschick1983formation}. The radio measurements suggest that the exciting stars are a ZAMS cluster of spectral types between O7.5 and B1.5.  From maps of the infrared fine-structure lines of  [NeII] 12.8\um and [Ar III] 8.99\um \citet*{beck1998infrared} estimate the spectral types to be O6-O7 and that each sub-source probably contains one exciting star.  We will suggest in this paper that the reality may be much more complicated.
\subsection{Observations}
A data cube  of the [NeII] 12.8\um (780.42 \wno) line was obtained at the NASA Infrared Telescope Facility (IRTF) on the night of 2009 July 13 with the University of Texas echelle spectrograph TEXES \citep{lacy2002texes}. TEXES is a high resolution spectrograph for wavelengths 5-25\um based on  a 256 $\times$ 256 element Raytheon Si:As array.  The diffraction and seeing limited beam size  on the IRTF is 1.4 arcsec~ and the spectral resolution, including thermal and instrumental effects, is $\sim 4 $\kms.  Each pixel along the N-S oriented slit was 0.36 arcsec; the slit was 1.37 arcsec~ wide and 9.3 arcsec~ long. The telescope was stepped $0.7 $ arcsec east-west across the source and subsequent scans separated by 5arcsec in declination; the ends of the scan were used for sky subtraction and the overlap to check and correct the matching of scan positions. No scans had to be shifted more than 2 arcsec  to match the next scan; pointing errors were $<1$arcsec. The scans were interpolated to create data cubes with pixels  0.6arcsec$\times$0.6 arcsec$\times$0.92\kms and 0.36 arcsec$\times$0.36  arcsec$\times$0.92\kms ~.  Both cubes are available online and the links given in the Appendix. This paper is based on the first cube, which samples each beam width twice.

\begin{figure*}
	\gridline{\fig{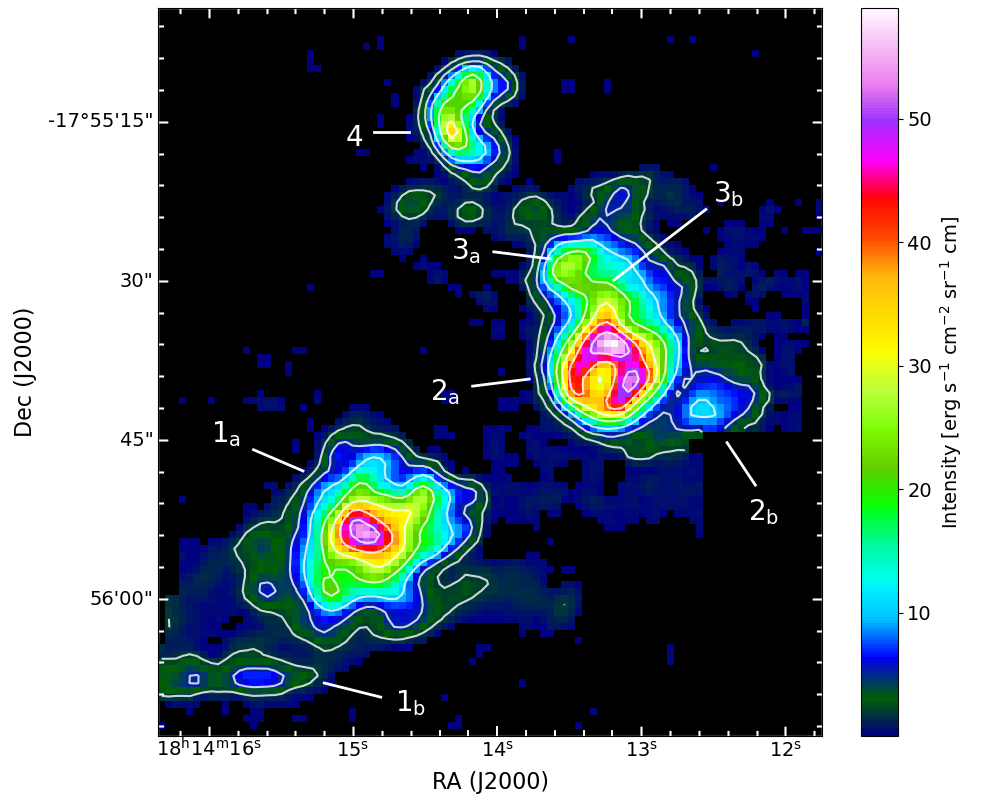}{0.5\textwidth}{(a)}
		\fig{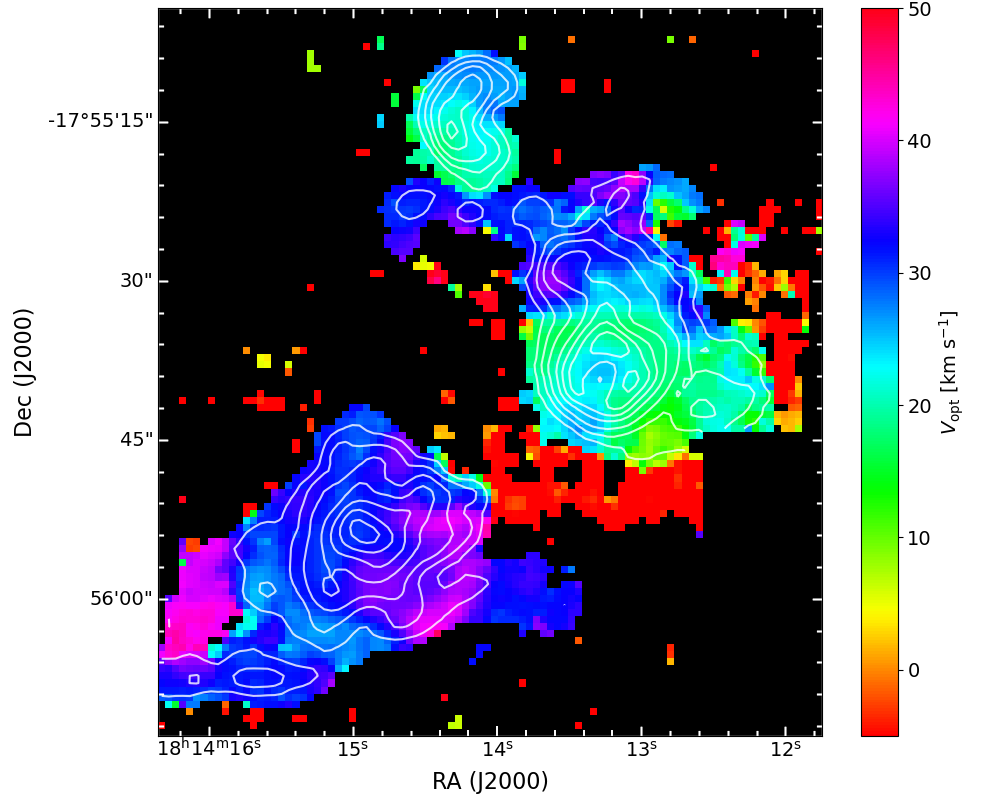}{0.5\textwidth}{(b)}}
	\caption{W33 Main [NeII] data cube (a) 0\textsuperscript{th} Moment Map showing the total [NeII] emission intensity, labelled with the names assigned the sub-sources. (b) 1\textsuperscript{st} moment of the data cube showing the weighted average velocity in color and the total intensity in contours.   The maps have $\sim$1.4 arcsec~ spatial resolution and  4 \kms kinematic resolution. Contour levels are [2,5,10,20,...,50] \flux.
		\label{fig:w33_main_0_1_moments}}
\end{figure*}

\begin{figure*}
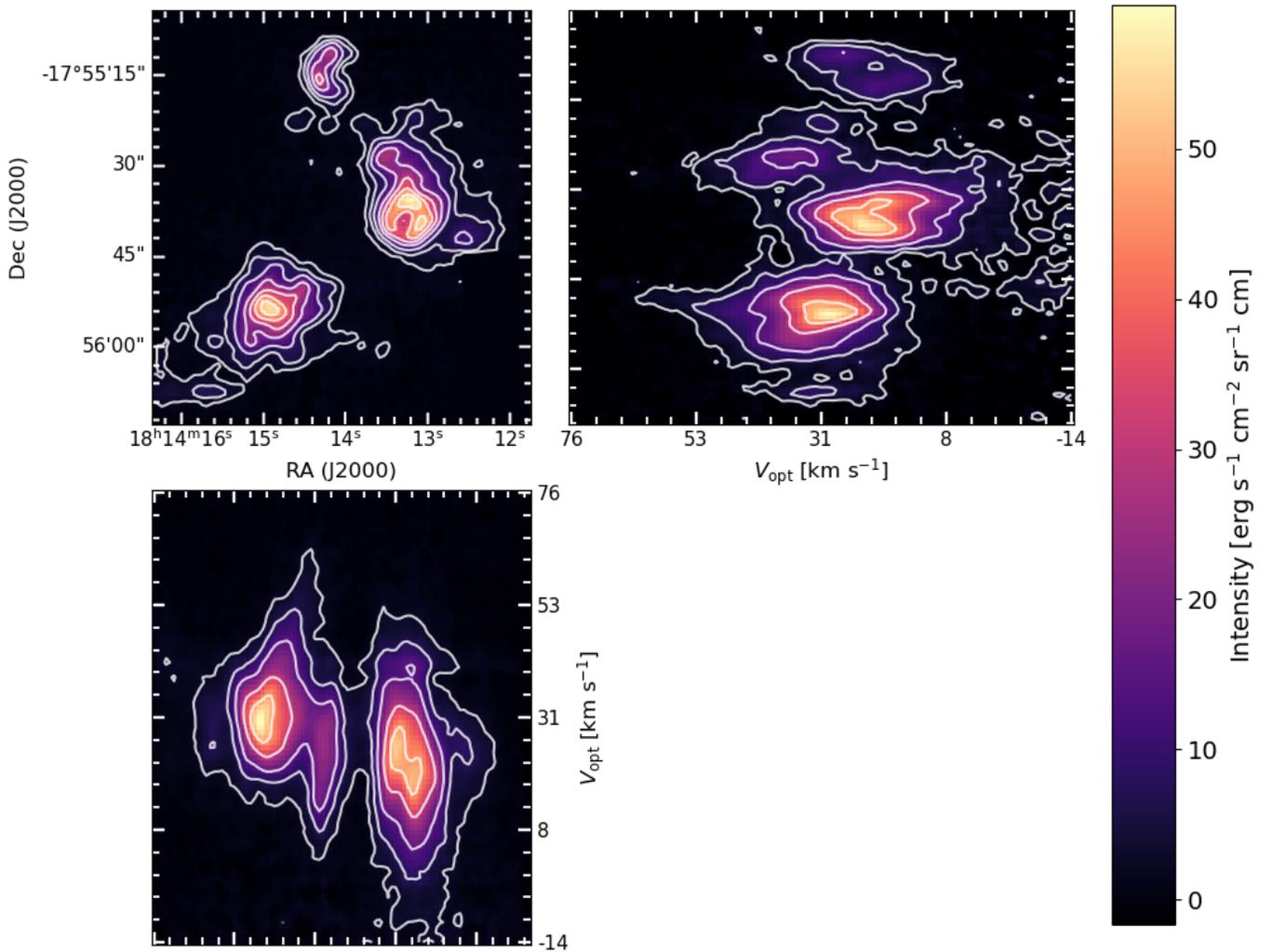

	\gridline{\fig{W33_Main_datacube.png}{1.0\textwidth}{}}
	\caption{W33 Main [NeII] data cube composite showing the 0\textsuperscript{th} moment maps along the spectral axis (top left), RA axis (top right) and Dec axis (bottom). The maps have 1.4 arcsec spatial resolution and 4\kms~ kinematic resolution. Contour levels are [2,5,10,20,30,40,50] \flux.
		\label{fig:w33_main_data cube}}
\end{figure*}

Atmospheric lines were used for wavelength calibration and a laboratory blackbody for flux calibration. We estimate the accuracy of the velocity registration to be $\pm1$\kms.~  The spatial position was registered by comparison to the radio interferometer maps and is estimated to be accurate to $\pm2$ arcsec.~  The data cube's 0\textsuperscript{th} and 1\textsuperscript{st} moment maps along the spectral axis are shown in Figure \ref{fig:w33_main_0_1_moments}, and Figure 
\ref{fig:w33_main_data cube} shows the 0\textsuperscript{th} moment maps along the two spatial axes, displaying the velocity extent of the emission in each source.  

Figures \ref{fig:w33_main_0_1_moments} and \ref{fig:w33_main_data cube} show clearly that the ionized gas is concentrated into at least three UCHII regions distinct in space and velocity.  Within each UCHII region the ionized gas has full velocity extent $\sim40$\kms;~ in addition the three UCHII regions are offset from each other in velocity by $\sim10$\kms, peaking at $\sim+33$ and $\sim+23$\kms.  The ionized gas integrated over the entire W33 Main region peaks at velocity $\sim+29$\kms.~  These UCHII regions have previously been detected in the IR and radio continuum, and the source identifications in the radio and infrared have not been consistent; the infrared sources were numbered from west to east as IRS1,2,3 and the radio continuum sources labelled as A,B,C from north to south.   We show on Figure \ref{fig:w33_main_0_1_moments} the labels we use throughout the paper, which  correspond to the infrared labels of \citet{beck1998infrared} as W33M1 (IRS 1), W33M2+3 (IRS 3) and W33M4 (IRS 2).   Note that the UCHII regions are not uniform but hold sub-structures and sub-sources that appear both in the intensity distribution and as velocity features; these are identified and labelled in the figure.   Figure \ref{fig:moment0_overlay} shows the total [NeII] emission map overlaid on the 6 cm radio map of \citet{haschick1983formation}.  This map shows that the ionized gas traced by [NeII] emission agrees with the radio continuum extremely well except for a diagonal band across the centre where there is radio continuum but no [NeII].  We believe this band to show where more \HII regions are so deeply embedded in the cloud that even the [NeII] is obscured; this feature is also seen in the Spitzer 8\um image of the region \citep{Churchwell_2009}.

\newpage
\section{Methodology} \label{sec:methodology}
\subsection{[NeII] as Tracer of Ionized Gas}
We use [NeII] as a tracer of the total ionized gas and assume that the ratio of [NeII] line intensity to the ionized gas emission measure is constant over the entire W33 Main region.   For this to be valid the gas-phase abundance of neon and the fraction of neon in Ne\textsuperscript{+} must be constant across W33 Main   and the gas densities well below the critical density  $n_c = 3.6 \times 10^5$ cm\textsuperscript{-3}  \citep*{Lacy_1982} at which quenching is significant; all these statements are consistent with previous work on W33 (constant neon abundance is discussed in \citep{beck1998infrared}).  We further assume that the extinction at 12.8\um is constant over the UCHII regions we model, consistent with the radio maps and the 8\um Spitzer  images.  
\subsection{Two Models of Cometary Structure in UCHII Regions} 
Two primary mechanisms for creating complex UCHII region structures are density gradients in the surrounding cloud and the motion of the exciting star through the gas.    \citet{Zhu_2015}  present hydrodynamical models showing that the 'cometary' structure observed for many UCHII regions can be created by a stationary star in an exponential density gradient.   Their model of 'champagne flow' is based on an exciting star which is stationary and close to the near face of the embedding cloud and predicts that gas in the 'tail' of the cometary structure will be blue-shifted relative to the head as the gas flows out of the cloud.  We will discuss this model further below in connection with W33M2,  the most  champagne-flow like of the sub-sources. The other sources in W33 Main do not show such clear cometary behavior; instead their appearance and kinematics suggest that exciting stars are moving through the cloud and creating bow shocks.  In the next section we  review bow-shock models and how we use them in this paper.

\subsubsection{ Bow Shock Models} \label{sec:bow_shock}

Bow shocks are of fundamental importance in UCHII regions and we here review how they will appear in the simulations.  When a  massive star with a stellar wind moves supersonically in a medium it creates a bow-shock \citep{wood1989morphologies,mac1991bow,van1992bow,mackey2015wind}:  the stellar wind and ambient material collide to form a shocked region in front of the star which is stationary in the frame of the star and approximately paraboloidal in shape \citep{wilkin1996exact}.  
The bow shock creates very dense neutral material that potentially traps the expanding ionization front and halts its expansion towards the front; this  over-pressure depends on the velocity of the star.  

Embedded stars moving through the dense ambient cloud are expected to create bow shocks.  Bow shocks can produce the cometary, core-halo, and shell structures which are the most common morphologies of UCHII regions, and prolong the lifetime of the UCHII stage.  The precise form taken depends on the ambient density, stellar velocity, nature of the stellar wind and the angle of observation; all these parameters need careful tuning to match the observations \citep{churchwell2002ultra}.

 \subsection{Hydrodynamics - PLUTO} \label{sec:pluto_hydrodynamics}
We created the models of the UCHII regions with the PLUTO hydrodynamics package \citep{mignone2007pluto}. All the UCHII region simulations were run inside a $256 \times 256 \times 256$ 3D Cartesian grid so that each grid cell was 0.0013 parsec across and the boundary conditions in all directions were set to outflow. The Navier-Stokes equations of classical fluid dynamics were solved in an Eulerian method in 3D Cartesian coordinates using a HLL approximate Riemann solver \citep*{harten1983upstream}. $T_e=10^4 K$ was assumed, heating and cooling were not included, and a Courant number of $0.3$ was used based on the Courant–Friedrichs–Lewy condition \citep*{courant1967partial}.

\begin{figure*}
	\gridline{\fig{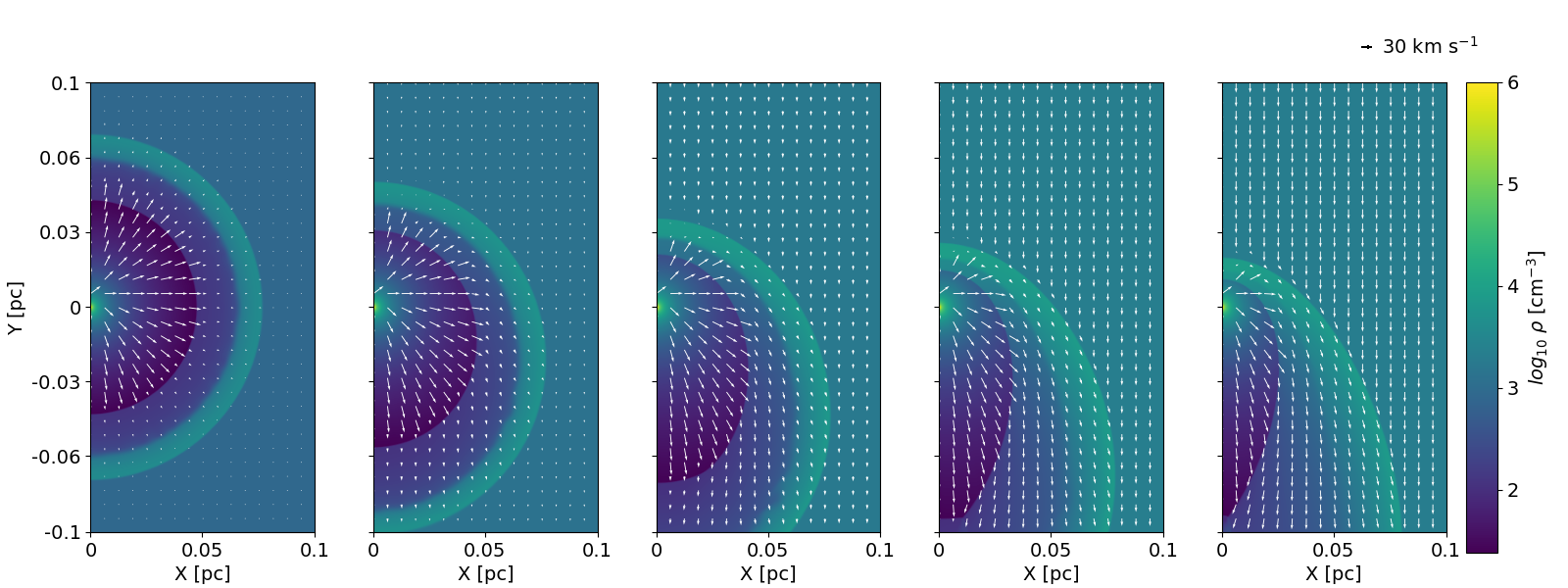}{1\textwidth}{}}
	\caption{Density plots, in axisymmetric cylindrical coordinates in the star's frame of reference and with the star at the origin,   of the cometary shape formed when the stellar wind interacts with the ambient material for a moving B0V star.   The cometary shape becomes more elongated with increasing stellar velocity from 0 \kms (far left) to 30 \kms (far right).
		\label{fig:cylindrical_stellar_wind}}
\end{figure*}

Each star starts with a spherically symmetric radial wind flowing from the origin of coordinates and moves through an ambient medium of constant density at stellar speed $v_*$ in the $\hat{j}$ direction. This gives initial conditions:
\begin{equation} 
	\setlength\arraycolsep{20pt}
	\begin{array}{ccc} 
		\rho = \rho\textsubscript{a} & p = p\textsubscript{a} & \pmb{v}=v_*\hat{j}
	\end{array} 
\end{equation}
The wind is injected within the internal boundary $r_0$ (inside the inner reverse shock of a stellar wind bubble) and based on \citet{mignone2014high} maintains these flow quantities constant in time:
\begin{equation} 
	\setlength\arraycolsep{20pt}
	\begin{array}{ccc} 
		r^2 v_r \rho = r_0^2 V_0 \rho_0 & v_r = V_0 \tanh\left(\frac{r}{r_0}\right) & p=\frac{c\textsubscript{s}^2}{\Gamma} (\rho_o^{1-\Gamma})\rho^\Gamma
	\end{array} 
\end{equation}
where $r$ is the radius, $\rho_0$ is the ambient density, $V_0$ is the gas velocity at $r_0$, $p$ is the gas pressure, c\textsubscript{s} is the sound speed of the gas and $\Gamma=c_P/c_V$ is the ratio of specific heat coefficients. We note that $r_0$ was set to a distance of between 3 and 6 cells from each star according to its appropriate mass, so that the effect of the stellar wind is captured at a sufficient resolution for our models; for a single star, increasing the resolution by a factor of 4 didn't change the various shock radii, the only major difference was that Rayleigh-Taylor instabilities developed at the forward shock radius because of the large density gradient \citep{meyer2014models}, however this would have a negligible effect on our results, especially for the interactions of multiple star trails, and therefore wouldn't affect the fine-tuning of the stellar parameters. We note that at much longer timescales these approximations can lead to significant discrepancies from analytical models of stellar wind bubbles \citep{Pittard2021}
With dimensions such that the spherical wind shell maintains $r_0 = V_0 = \rho_0 = 1$ the effects of  increasing stellar velocity are shown in Figure \ref{fig:cylindrical_stellar_wind}.  The stationary star creates a Strömgren sphere but forms a cometary tail when  moving supersonically. The faster the star moves,  the more elongated the cometary shape and the greater the ram pressure at its apex. 

\begin{figure*}
	\gridline{
		\fig{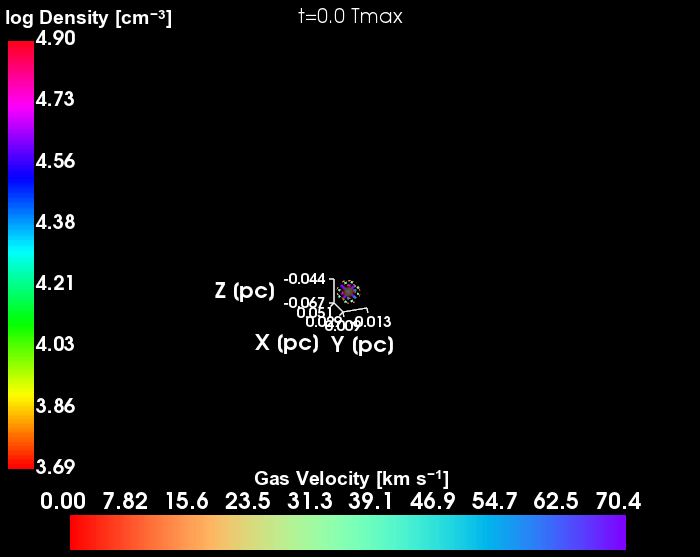}{0.35\textwidth}{(a)}
		\fig{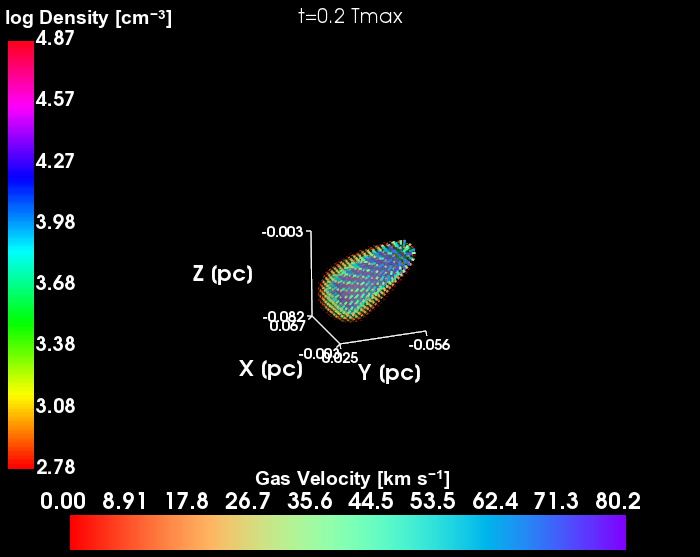}{0.35\textwidth}{(b)}
		\fig{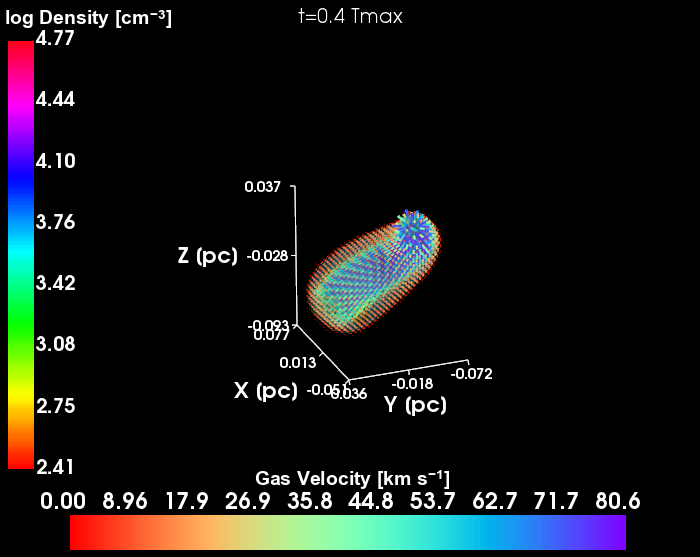}{0.35\textwidth}{(c)}
	}
	\gridline{
		\fig{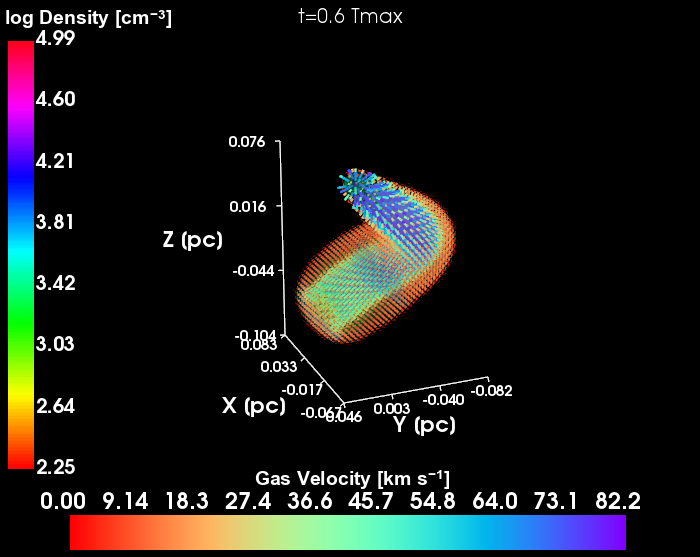}{0.35\textwidth}{(d)}
		\fig{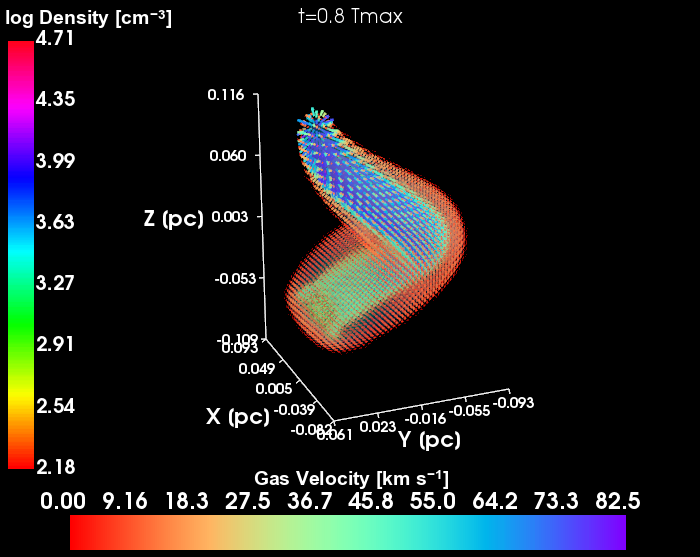}{0.35\textwidth}{(e)}
		\fig{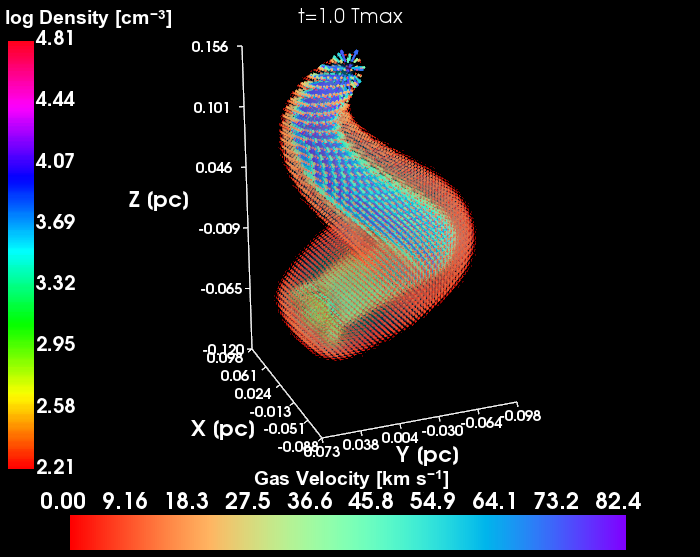}{0.35\textwidth}{(f)}
	}
	\caption{Example of the curled tail that is left behind, at multiple timesteps, by a B0V star performing a complete orbit and moving at 20\kms in the Z+ axis direction with orbital parameters perpendicular to the Z axis of: $r\textsubscript{orb}=0.040$ pc, $\omega\textsubscript{orb}=$ $2.03\times10^{-11}$ and $\varphi\textsubscript{orb}=0$ rad. \label{fig:curled_tail_example}}
\end{figure*}

The trail remains imprinted on the medium for some time tracing the star's movement, and when the stellar trajectory is not a straight line the result is  a short curled tail.  An example of the curled tail left behind by the circular motion of a B0V star moving at 20\kms~ can be seen in Figure \ref{fig:curled_tail_example}.   Similar structures are seen in density maps from simulations of the interaction between the stellar winds of two stars \citep{van2019simulating}, and observed in the environment of the evolved star Mira \citep{martin2007}. 
The similarity to the observed data cube motivated us to model the W33 UCHII regions by combining multiple stars, each with a wind and curved trajectory. 
 
We assumed that the stellar orbits are circular in $x,y$ and that within one UCHII region the stellar orbits are co-planar. Therefore, each star in orbital motion moves on a  curve described by
\begin{equation} 
	\setlength\arraycolsep{20pt}
	\begin{array}{ccc} 
		x = x_0 + r\textsubscript{orb} \cos(\omega\textsubscript{orb}t + \varphi\textsubscript{orb}) & y = y_0 + r\textsubscript{orb}\sin(\omega\textsubscript{orb}t + \varphi\textsubscript{orb}) & z = z_0 + z\textsubscript{vel} t
	\end{array} 
\end{equation}
where ($x_0,y_0,z_0$) is initial position of each star, $z\textsubscript{vel}$ is the stellar velocity and $r\textsubscript{orb}$, $\omega\textsubscript{orb}$ and $\varphi\textsubscript{orb}$ are respectively the orbit's radius, circular velocity and phase.

We wrote a 3D visualization program that allowed us to modify wind, orbit and stellar parameters for each star so as to duplicate the morphology of each UCHII region in the data cube.  The adjustable parameters and the best value we found for every star are shown in Table \ref{tab:pluto_stars_parameters}.  In Figures \ref{fig:w33_1_star_curves}, \ref{fig:w33_2+3_star_curves} and \ref{fig:w33_4_star_curves} we show 3-dimensional views of the best model stars and their trajectories,  as well as of the original data cube. 
As part of the fine-tuning process we determined the ratio between the inner reverse shock radius and the radius of the shocked ambient gas, and used this with the formulae of   \citet{weaver1977interstellar} to derive the spectral type, mass and temperature of each star.  We assumed  OB star values of \citep{lamers1993mass,krtivcka2014mass} and that the UCHII region had been expanding for  \tilded $10^3-10^4$yr.   

\begin{deluxetable}{ccccccccccc}
	\tablenum{1}
	\tablecaption{PLUTO Stars and Movement Parameters \label{tab:pluto_stars_parameters}}
	\tablewidth{0pt}
	\tablehead{
		\colhead{UCHII Region} & \colhead{Star \#} & \colhead{Spectral Type} & \colhead{$M_* $[M\textsubscript{\(\odot\)}]} & \colhead{$T_*\textsuperscript{eff}$ [K]} & \colhead{$v_*$ [\kms]} & \colhead{$r_0$ [$10^{-3}$ pc]} & \colhead{$V_0$ [\kms]} & \colhead{$r\textsubscript{orb}$ [pc]} & \colhead{$\omega\textsubscript{orb}$ [$10^{-12}$ s\textsuperscript{-1}]} & \colhead{$\varphi\textsubscript{orb}$ [rad]}
	}
	\startdata
	W33 Main 1\textsubscript{a} & 1* & B1.5V & 9.26 & 24000 & 24.12 & 4.26 & 40 & 0.014 & 6.75 & 1.05 \\
	- & 2 & B0.5V & 12.69 & 28000 & 21.93 & 6.39 & 60 & 0.021 & 9.33 & 4.97 \\
	- & 3 & B0.5V & 12.69 & 28000 & 21.93 & 6.39 & 60 & 0.020 & 7.58 & 1.05 \\
	- & 4* & B1.5V & 9.26 & 24000 & 12.06 & 4.26 & 40 & 0.013 & 7.65 & 0.05 \\
	- & 5* & B1V & 10.88 & 26000 & 17.54 & 4.79 & 45 & 0.007 & 11.26 & 3.12 \\
	- & 6 & B0.5V & 12.69 & 28000 & 19.74 & 6.39 & 60 & 0.016 & 9.33 & 2.48 \\
	- & 7* & B1V & 10.88 & 26000 & 17.54 & 4.79 & 45 & 0.021 & 7.00 & 0.61 \\
	W33 Main 1\textsubscript{b} & 1 & B1.5V & 9.26 & 24000 & 5.48 & 3.19 & 30 & 0.011 & 2.10 & 3.14 \\
	W33 Main 2\textsubscript{a} & 1 & B1V & 10.88 & 26000 & 21.93 & 5.32 & 50 & 0.015 & 10.12 & 5.29 \\
	- & 2 & B1V & 10.88 & 26000 & 19.74 & 5.32 & 50 & 0.021 & 9.33 & 0.87 \\
	- & 3 & B1V & 10.88 & 26000 & 21.93 & 4.79 & 45 & 0.037 & 9.33 & 3.21 \\
	- & 4 & B1.5V & 9.26 & 24000 & 19.74 & 4.26 & 40 & 0.018 & 9.33 & 6.16 \\
	- & 5 & B1V & 10.88 & 26000 & 15.35 & 4.79 & 45 & 0.016 & 5.83 & 0.38 \\
	W33 Main 2\textsubscript{b} & 1 & B1.5V & 9.26 & 24000 & 4.24 & 4.26 & 40 & - & - & - \\
	W33 Main 3\textsubscript{a} & 1 & B1.5V & 9.26 & 24000 & 8.22 & 3.19 & 30 & - & - & - \\
	W33 Main 3\textsubscript{b} & 1 & B1.5V & 9.26 & 24000 & 5.48 & 3.72 & 35 & - & - & - \\
	W33 Main $4$ & 1 & B1.5V & 9.26 & 24000 & 7.89 & 4.26 & 40 & 0.021 & 5.83 & 0.00 \\
	- & 2 & B1.5V & 9.26 & 24000 & 10.53 & 3.72 & 35 & 0.005 & 19.83 & 0.52 \\
	\enddata
	\tablecomments{	List of stars simulated within each UCHII region. Stars marked with '*' are potentially simple gas clumps. For each star its listed parameters include mass ($M_*$), effective temperature ($T_*\textsuperscript{eff}$), velocity relative to the ambient material ($v_*$), internal boundary of its stellar wind bubble ($r_0$), $V_0$ is the gas velocity at $r_0$, orbital movement radius ($r\textsubscript{orb}$), angular velocity ($\omega\textsubscript{orb}$) and phase angle ($\varphi\textsubscript{orb}$). Note that these orbital parameters do not describe the physical orbit but refer to the circular paths of the simulations which approximate the star's relative movement.
	}
\end{deluxetable}
\subsection{3D Ionic Line Emission Profile Simulation - RADMC-3D} \label{sec:radmc3d_3d_profile} 
The PLUTO outputs (Ne\textsuperscript{+} density, velocity and position) were input to the radiative transfer software RADMC-3D  \citep{dullemond2012radmc} which  calculated the 3D [NeII] line profile emission.  We assumed constant gas temperature $10^4$K.   The line profile for each UCHII region in W33 Main was simulated separately at the spectral range and resolution matching the observed [NeII] data cube.  However the zero point of the velocity scale in the simulations was chosen for computational convenience,  and is not the velocity of the W33 cloud, so the velocities displayed for the simulation results are relative only.  When the original data is presented, the velocities are the observed values. 
\section{Simulation Results} \label{sec:results}

In this section we show the results of the 3D hydrodynamic and line profile simulations, fine tuned to match each UCHII region.  Since the simulations are on a  0.6 arcsec~  grid we convolve the results with a Gaussian to create an effective PSF (point spread function) of 1.4 arcsec, matching that of the observations, and compare them to the original data cube. 
We now present, for each sub-source, the original data and the locations and trajectories of the stars which produce the best fit.  We then compare the kinematics and structure of the simulated model and the original data in the form of channel maps and position-velocity diagrams (PVDs).  In section \ref{sec:results_population} we discuss and check the results in the context of star formation models.
\subsection{W33 Main 1} \label{sec:results_w33_main_1}

\begin{figure*}
\gridline{\fig{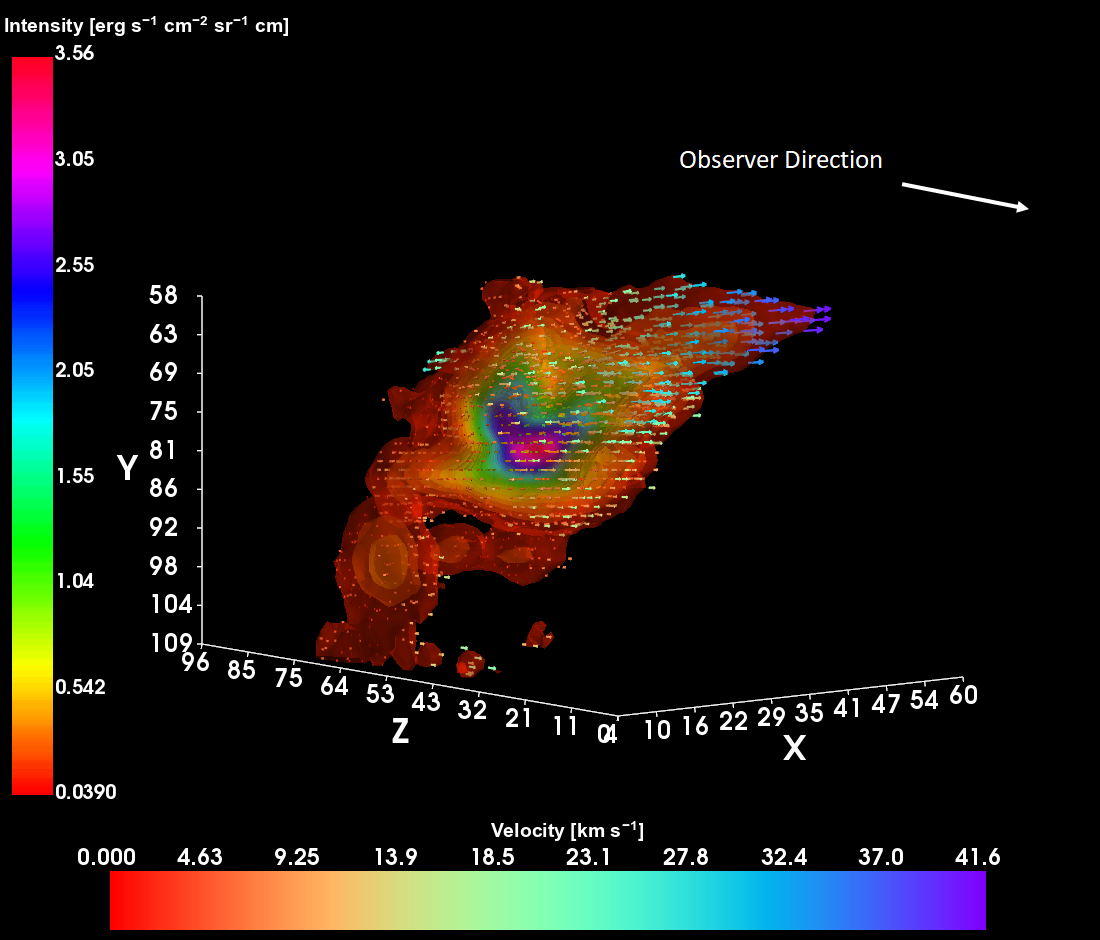}{0.37\textwidth}{(a)} 
       \fig{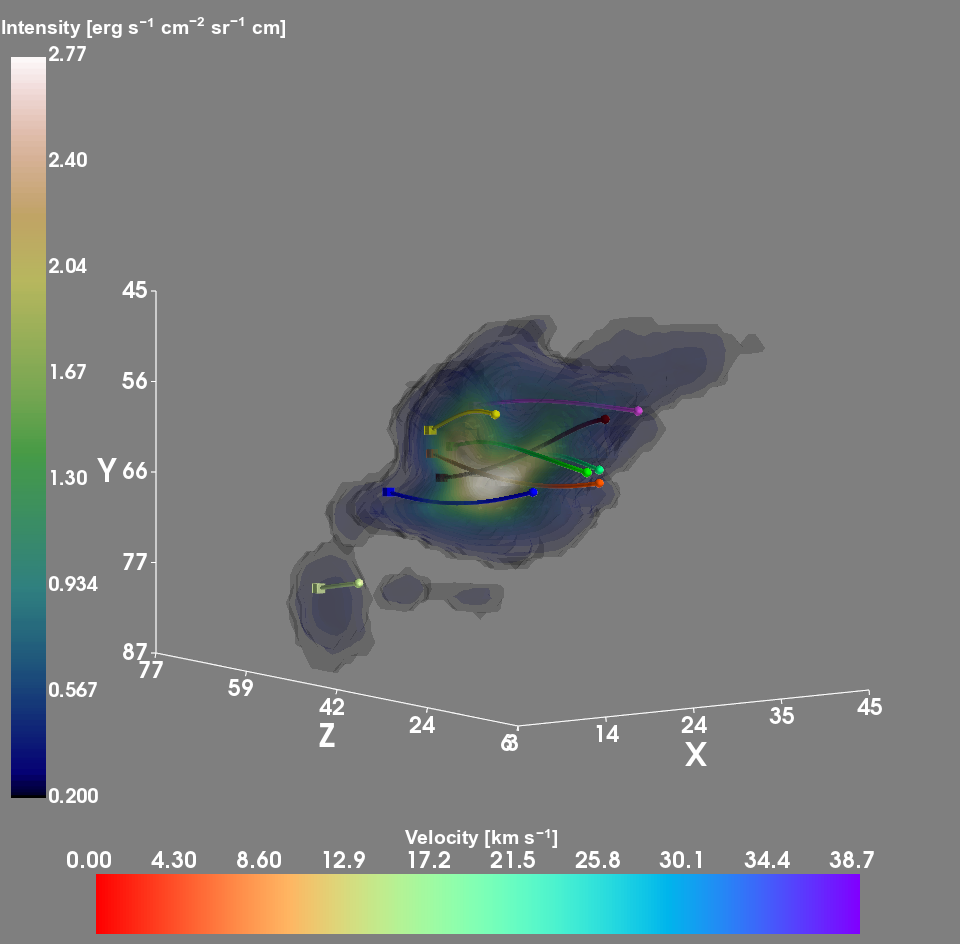}{0.37\textwidth}{(b)}}
\gridline{\fig{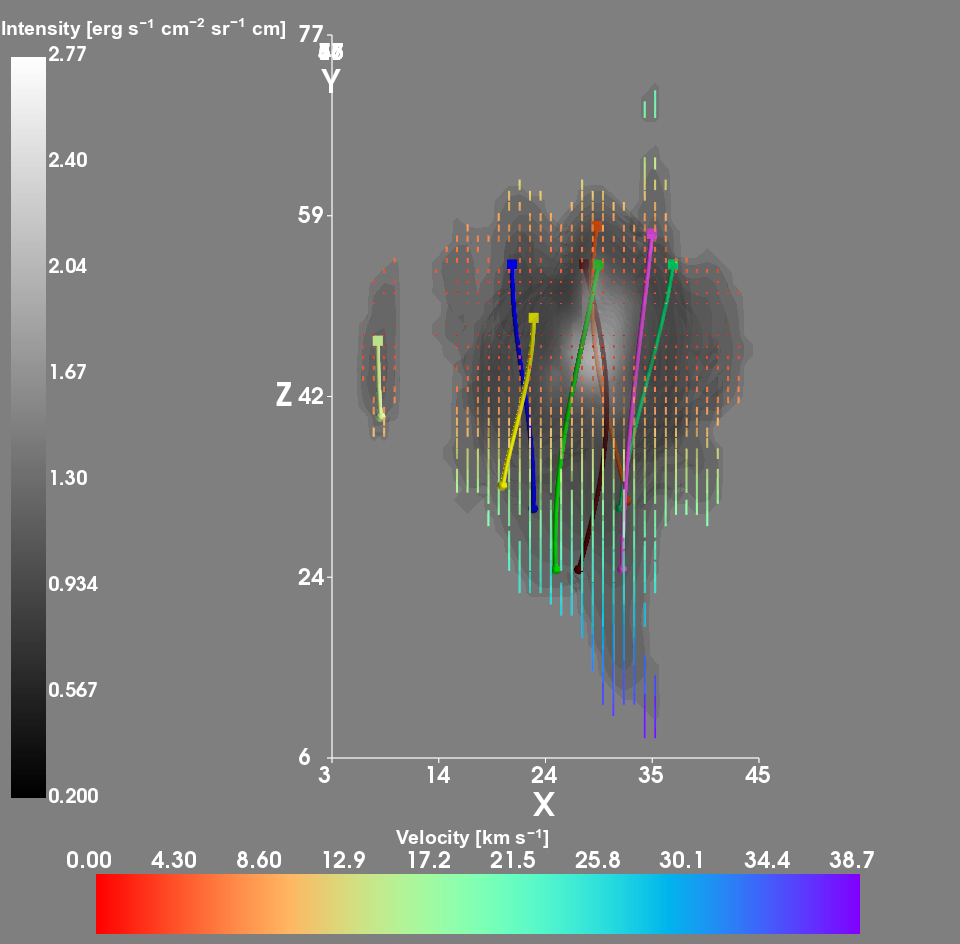}{0.37\textwidth}{(c)}
		\fig{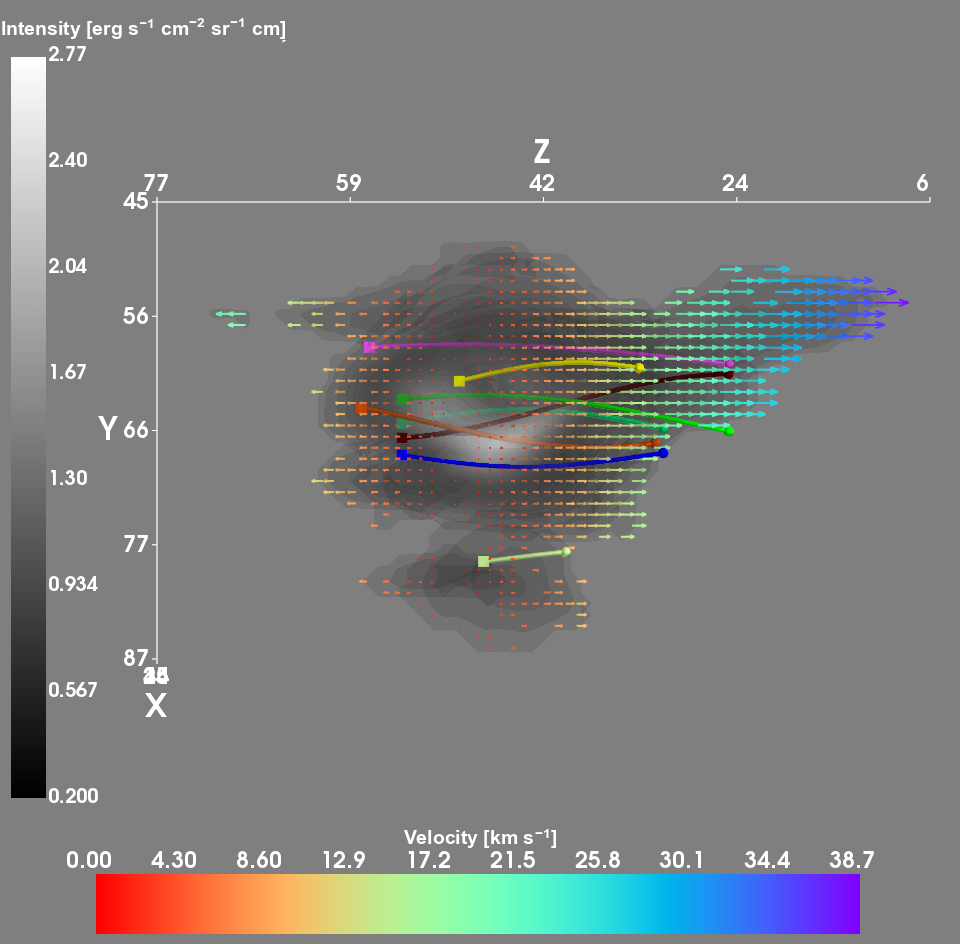}{0.37\textwidth}{(d)}}
\gridline{\fig{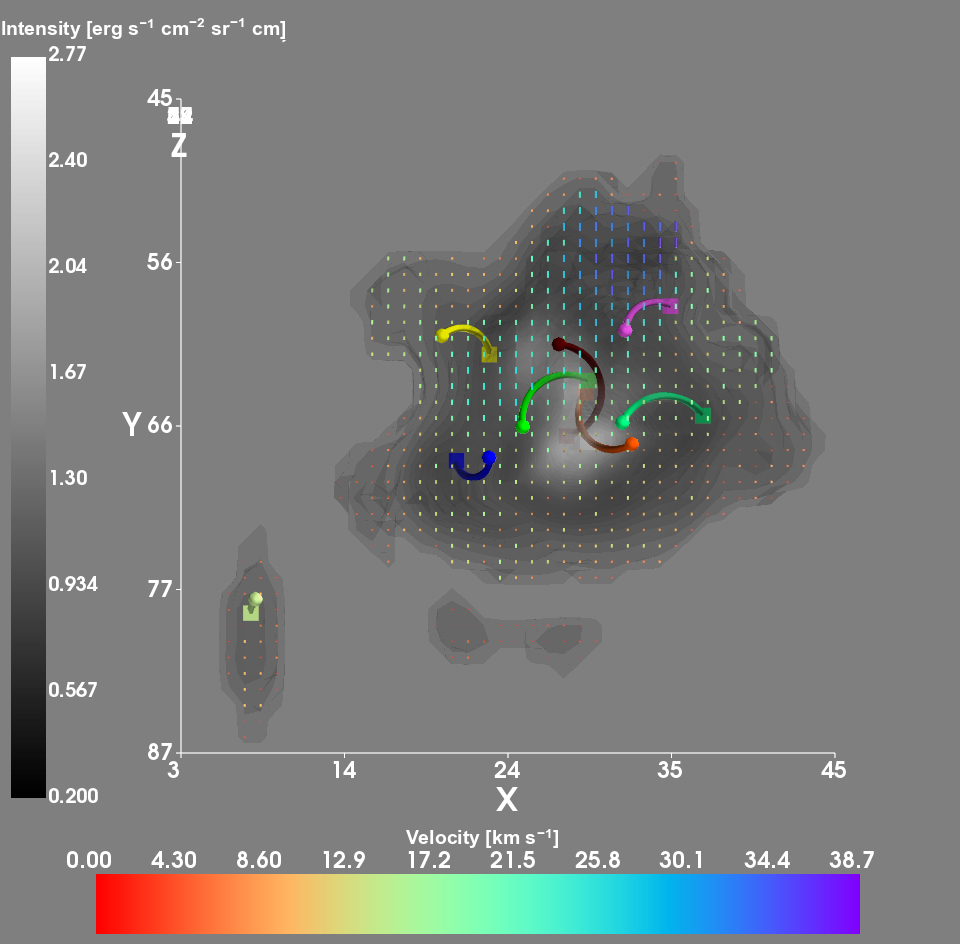}{0.37\textwidth}{(e)}
        \fig{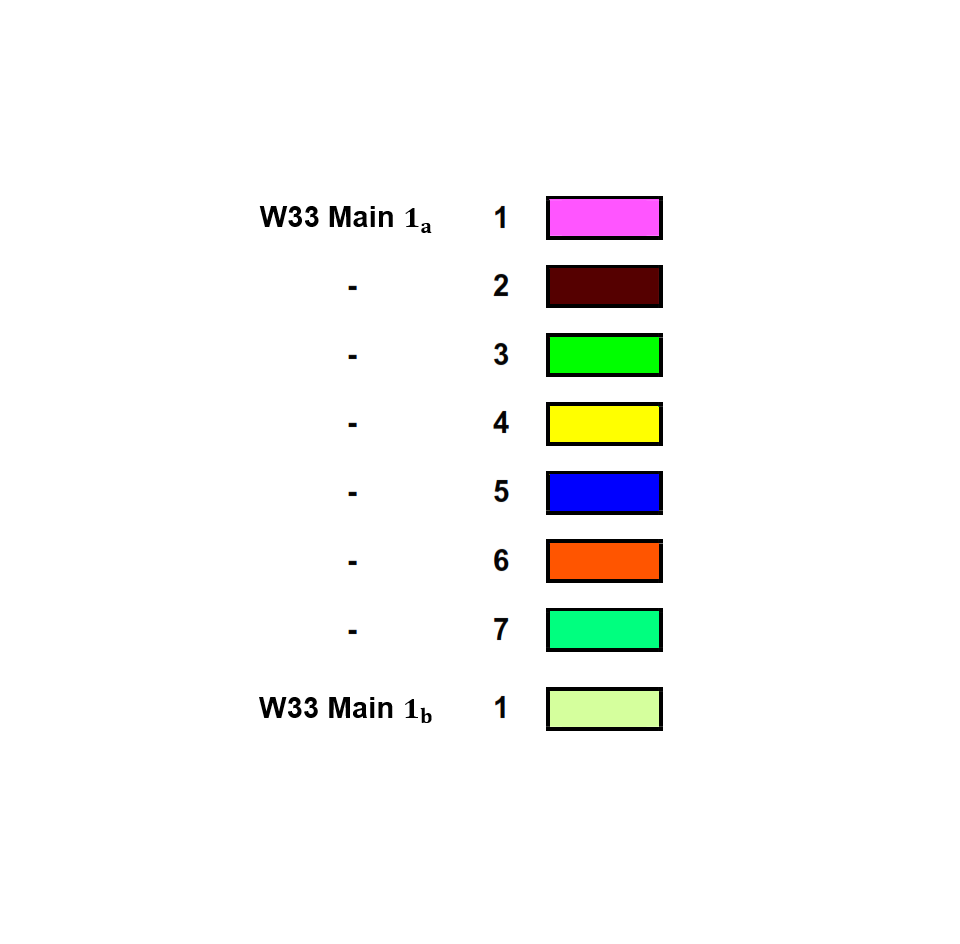}{0.37\textwidth}{}}
		\caption{3D View of the original [NeII] data cube cropped to the volume containing W33 Main 1 with the colored star movement curves (legend on bottom right) from 3 axes directions (Z- is the observer direction): (a) Dec, (b) RA, (c) Spectral/Kinematic. The X,Y,Z units are as follows: X axis is RA (1 X Unit = 0.66 arcsec), Y axis is Dec (1 Y Unit = 0.05s) and Z axis is spectral/kinematic axis (1 Z Unit = 1 \kms). The velocity axis is also denoted by the colored arrows where only the absolute velocity is considered. Each star curve shows the movement of a single star starting from the sphere shape and ending with the cube shape.}
		\label{fig:w33_1_star_curves}
\end{figure*}

W33M1 (W33 Main 1) includes W33M1\textsubscript{a} which has an irregular morphology and W33M1\textsubscript{b} which is much smaller and has a cometary form. The clues for the trajectories of stars and ionized gas clumps in W33M1\textsubscript{a} are subtle. Using our visualization program to fine-tune the models we have determined the  extent and shape of each separate inner star clump and display the curve of its orbit.   The simulations  best fitting the data, shown in Figure \ref{fig:w33_1_star_curves}, have W33M1\textsubscript{b} as a single star and W33M1\textsubscript{a} with as many as 7; 3 stars closer to the centre and the rest in the outer parts.  Note that some of these may be hot dense gas clumps which are distinct in the data cube but may not hold massive stars. 

\begin{figure*}[t]
	\gridline{\fig{W33M1_Orig_PVD_Dec.png}{0.54\textwidth}{(e)}
		\fig{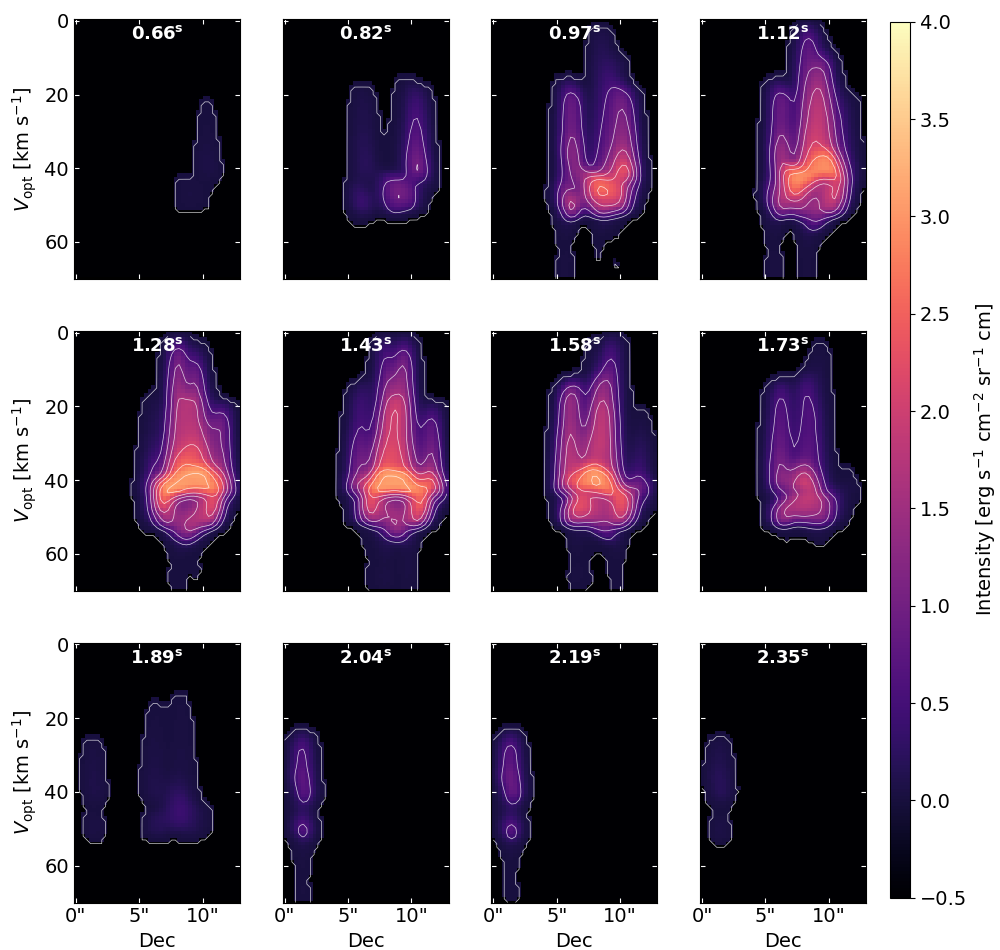}{0.54\textwidth}{(f)}}
	\caption{\scriptsize Comparison of original data cube and the simulation results for W33 Main 1: Channel maps of (a) the W33 Main [NeII] data cube cropped with contour levels at [0.05, 0.1, 0.2, 0.4, 0.6, 0.8, 1.0, 1.5, 2, 2.5, 3] \flux~  and (b) line profile emission simulation results with contour levels at [0.0, 0.5, 1, 1.5, 2, 2.5, 3.0, 3.3, 3.5, 3.7, 4.0] \flux. (c) Position-Velocity Diagrams of W33 Main 1 in RA from the original data and (d) from the simulation. (e) Position-Velocity Diagrams of W33 Main 1 in Dec from the original data and (f) the simulations.} 
	\label{fig:w33_main_1_results}
\end{figure*}

Figure \ref{fig:w33_main_1_results} compares the original data and model results for  W33M1. The channel maps of the simulation results agree with the original data in overall structure: there is a central peak surrounded by smaller clumps. The simulations have a more organized appearance; the secondary clumps are stronger and more defined than in the original data.     The  PVDs taken from the data and the simulations also agree on large scales, with a concentrated central source and a spatially extended high velocity 'tail' in both RA and Dec.  The PVDs from the simulated data have more pronounced small-scale structure: specifically, the high velocity tail in the simulations is clearly split into two separate spatial branches, a feature which is present but much less clearly so in the original data.   The tendency of the  secondary features to be more pronounced in the simulations than in the original data suggests that some non-stellar density structures may have been modelled as OB stars.   We note that  the morphology of W33M1\textsubscript{a} is not easily reproduced.  Further fine-tuning of the stellar motion curves may improve the correspondence.  

\begin{figure*}[htp]
	\gridline{\fig{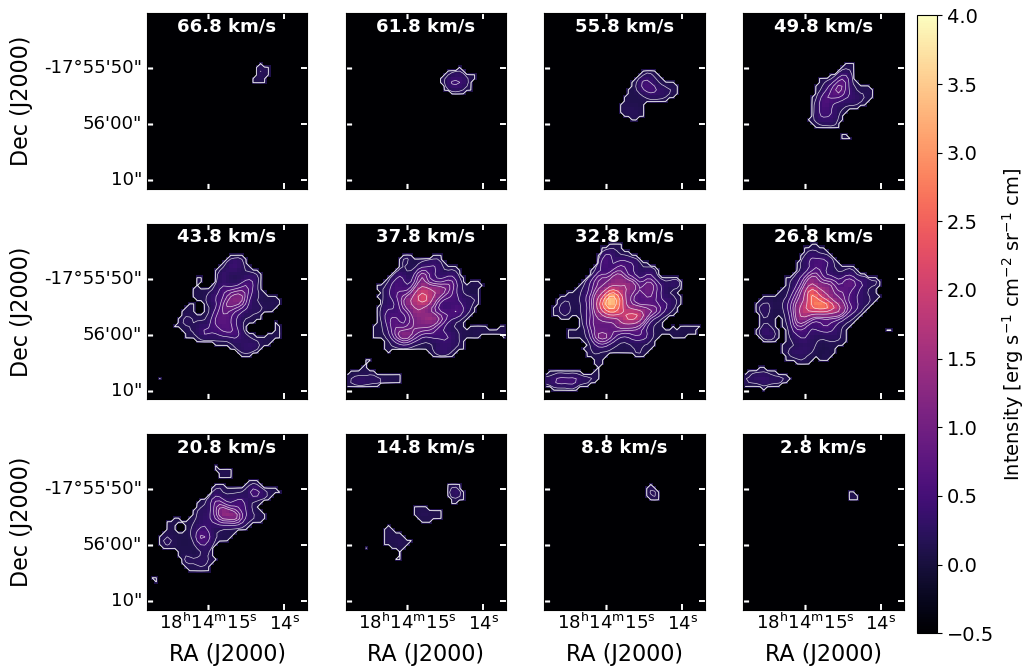}{0.54\textwidth}{(a)}
		\fig{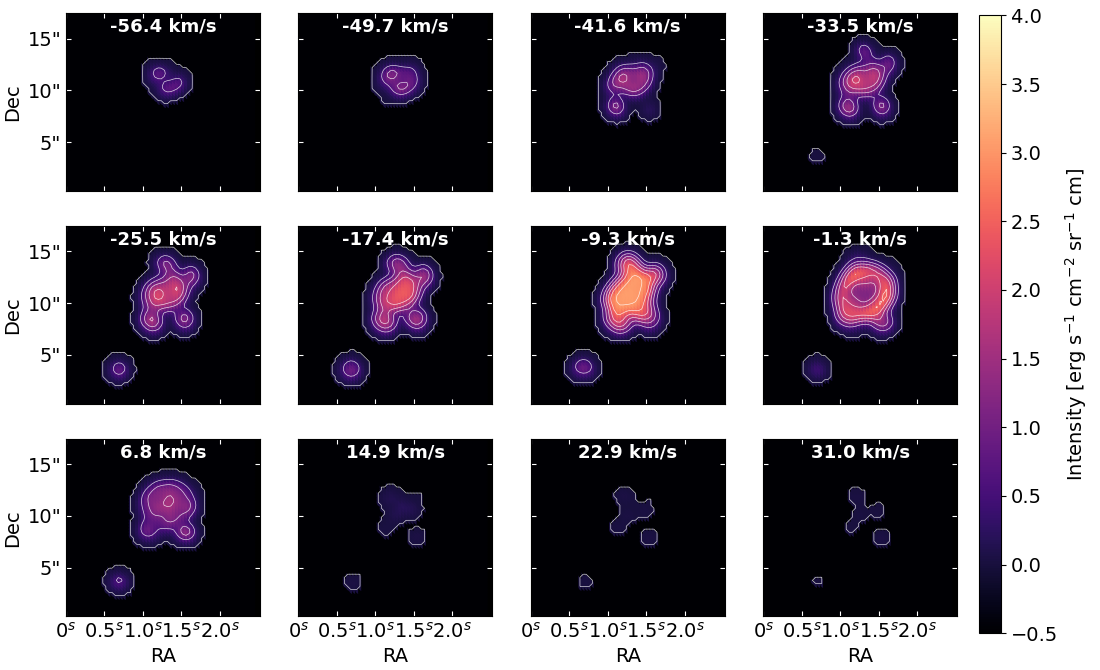}{0.54\textwidth}{(b)}}
\gridline{\fig{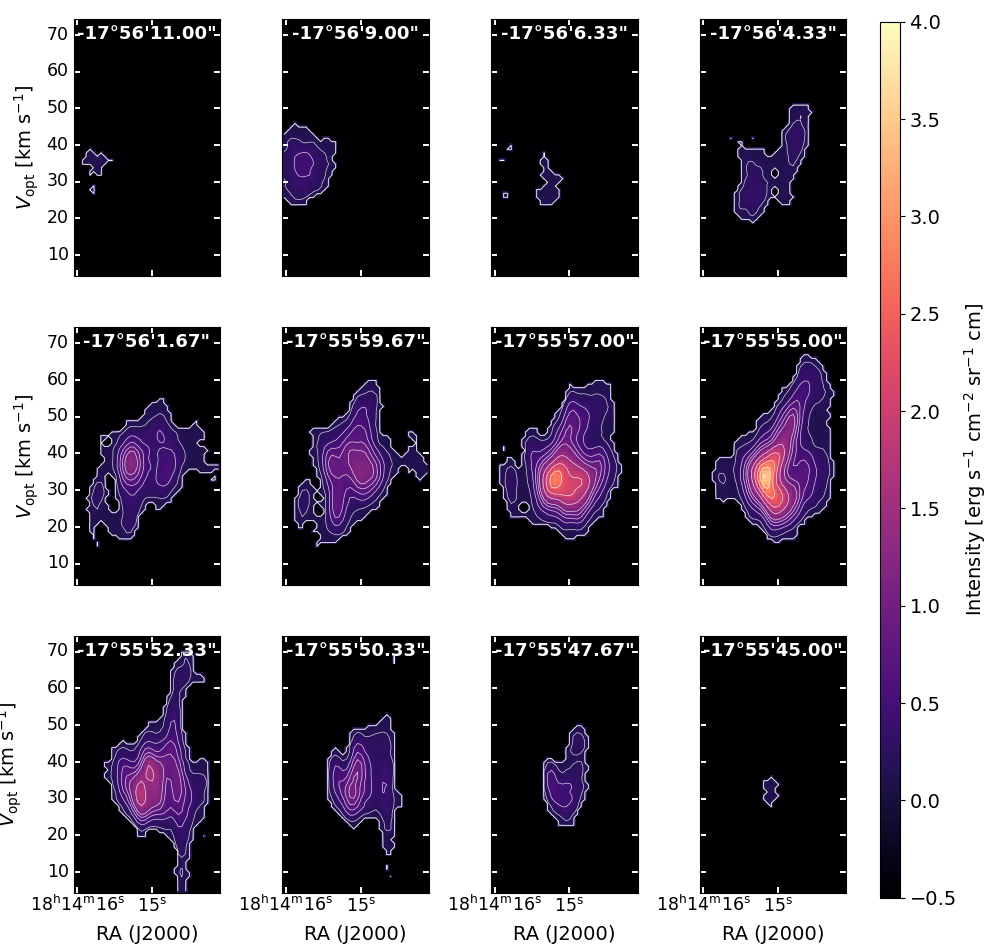}{0.54\textwidth}{(c)}
		\fig{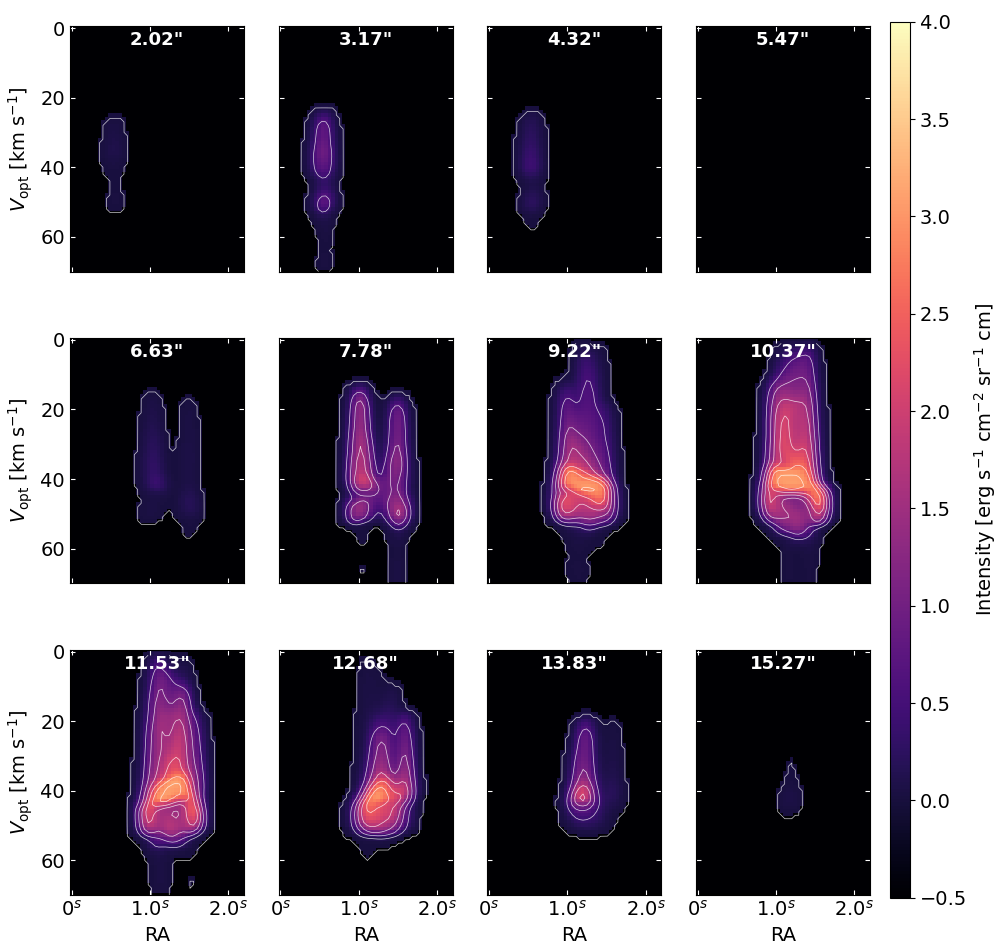}{0.54\textwidth}{(d)}}
		\vspace*{5cm}
\end{figure*}

\subsection{W33 Main 2 + 3} \label{sec:results_w33_main_2-3}

\begin{figure*}
\gridline{\fig{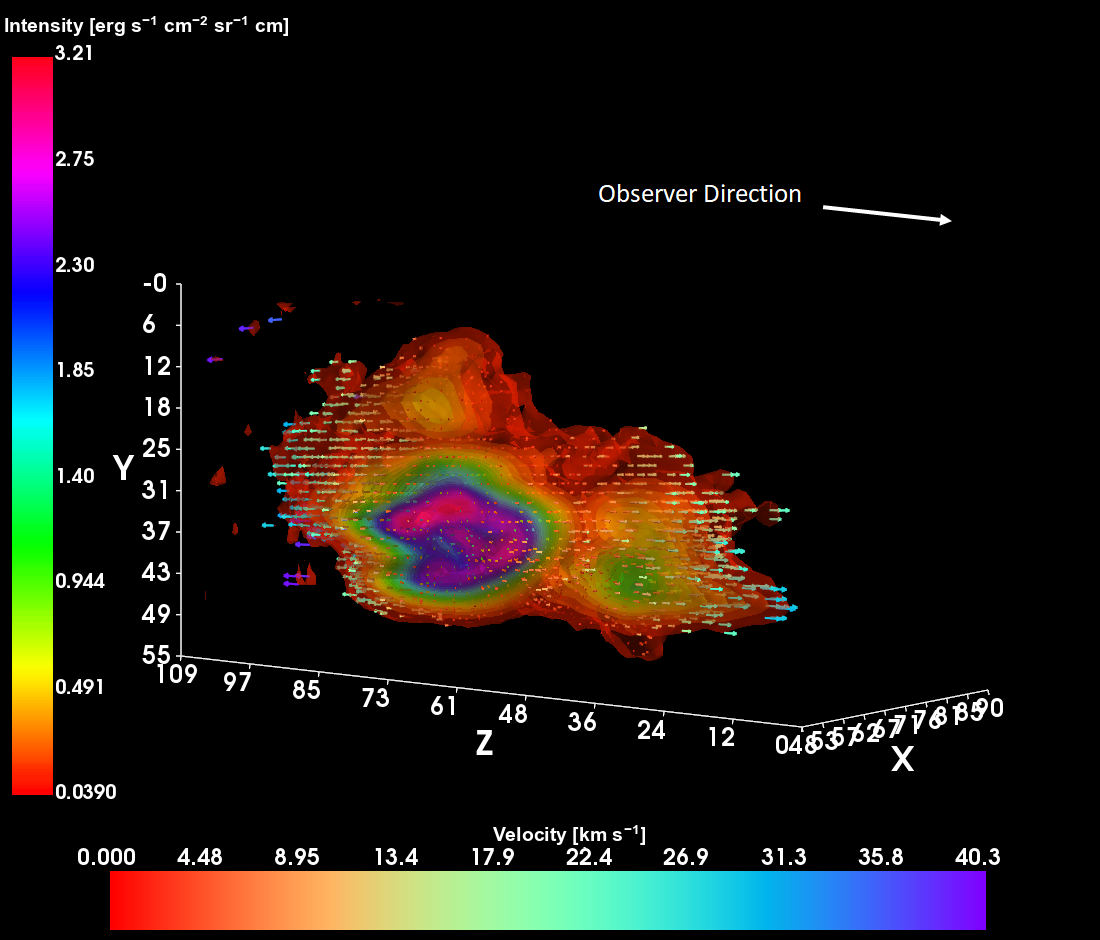}{0.37\textwidth}{(a)}
		\fig{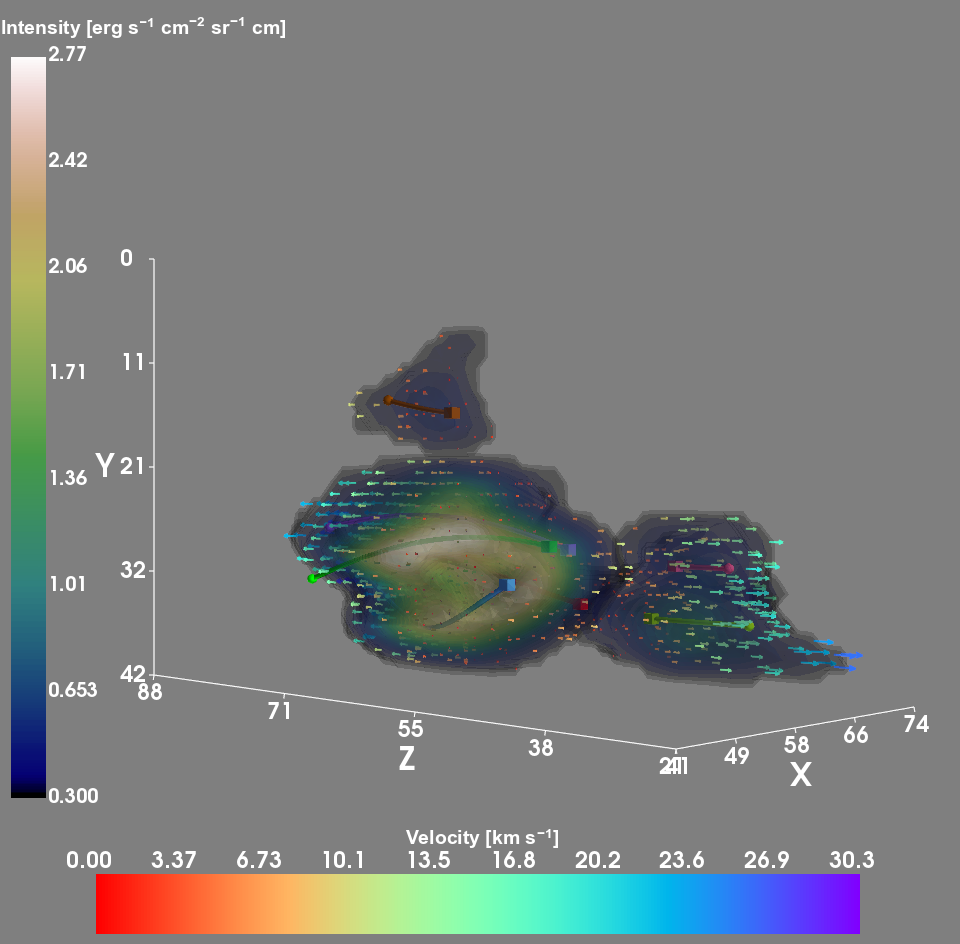}{0.37\textwidth}{(b)}}
\gridline{\fig{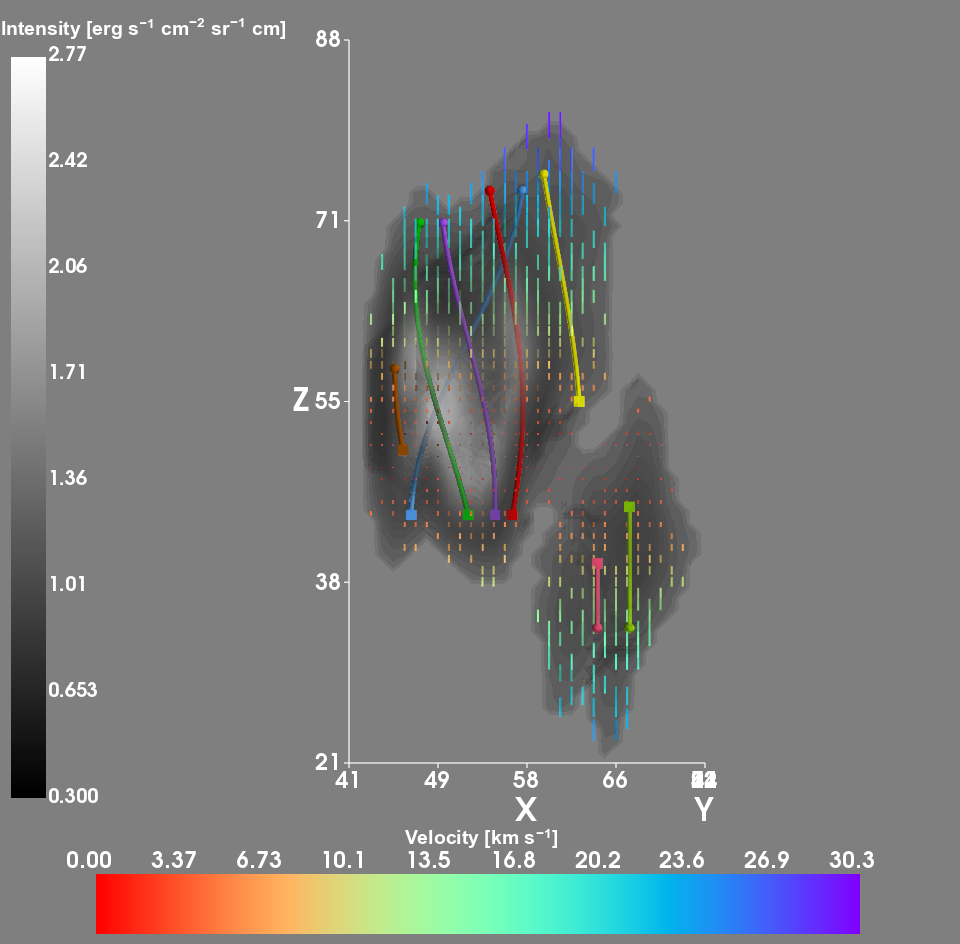}{0.37\textwidth}{(c)}
		\fig{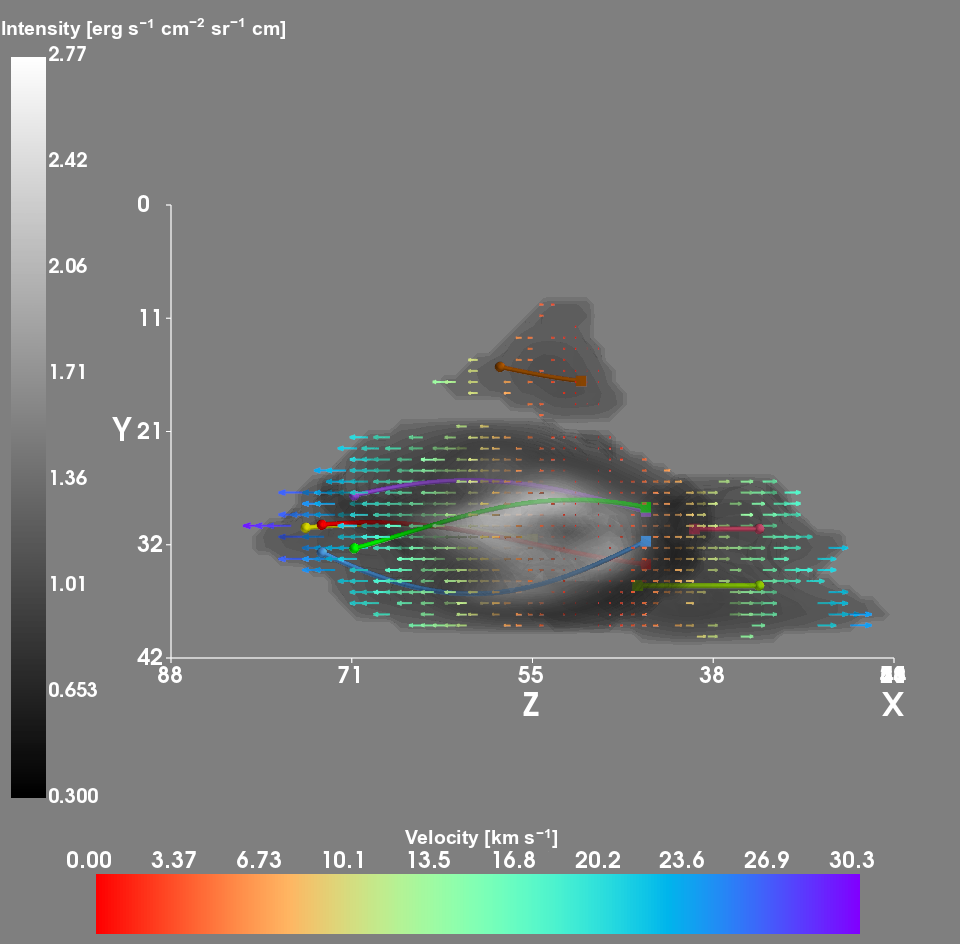}{0.37\textwidth}{(d)}}
	\gridline{\fig{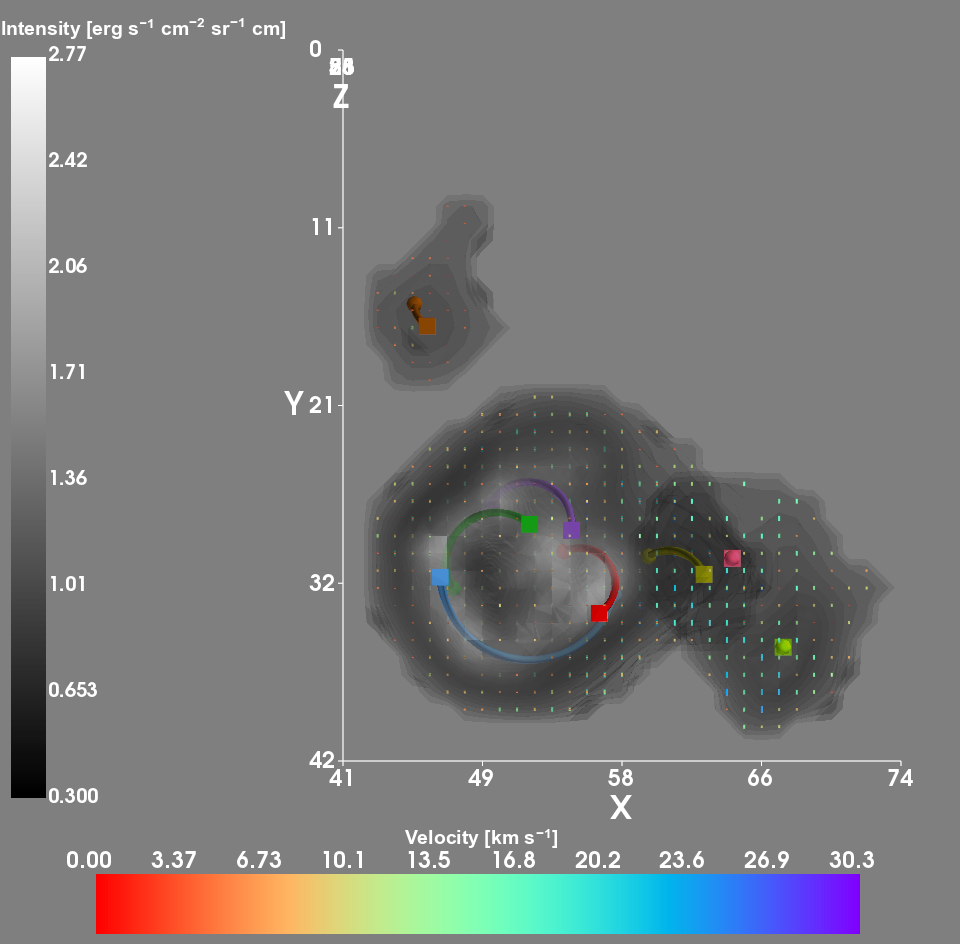}{0.37\textwidth}{(e)}
		\fig{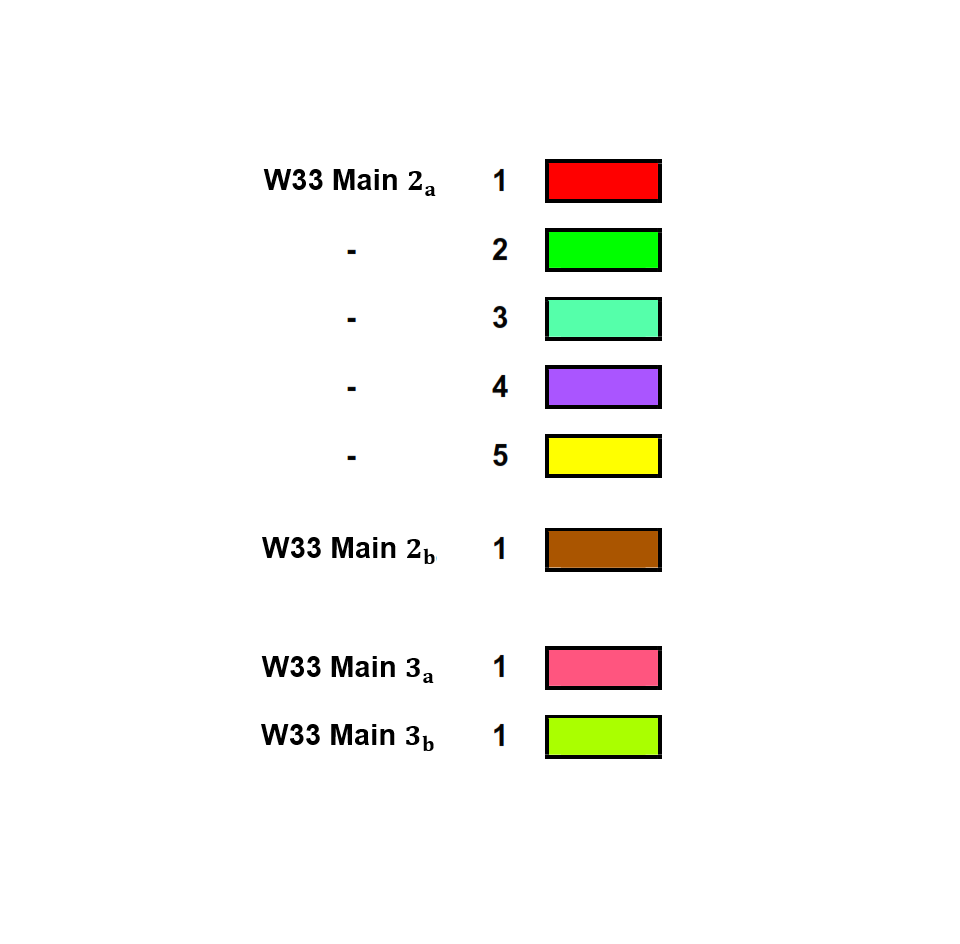}{0.37\textwidth}{}}
		\caption{3D View of the original [NeII] data cube cropped to the volume containing W33 Main 2+3  with the colored star movement curves (legend on bottom right) from 3 axes directions (Z- is the observer direction): (a) Dec, (b) RA, (c) Spectral/Kinematic. The X,Y,Z units are as follows: X axis is RA (1 X Unit = 0.66 arcsec), Y axis is Dec (1 Y Unit = 0.05s) and Z axis is spectral/kinematic axis (1 Z Unit = 1 \kms). The velocity axis is also denoted by the colored arrows where only the absolute velocity is considered. Each star curve shows the movement of a single star starting from the sphere shape and ending with the cube shape.}
		\label{fig:w33_2+3_star_curves}
\end{figure*}

Examination of the data cube shows that W33M2 and W33M3 are moving in opposing directions and grazing each other in a partial collision. We therefore treat them together in our simulations.   Figure \ref{fig:w33_2+3_star_curves} shows that W33M2\textsubscript{a}, W33M2\textsubscript{b}, W33M3\textsubscript{a} and W33M3\textsubscript{b} all appear somewhat cometary in morphology.  W33M2\textsubscript{a} is particularly clear because it appears that our observations look directly into the tail, and that the tail gas is expanding out of the cloud towards us, as shown by the blue-shift of the tail relative to the rest of the cloud.    W33M2\textsubscript{a} may fit the picture of a pure champagne flow UCHII region and this model is presented in the Discussion section on this source.    
We here treat both  W33M3, which does not have the same kinematic signature of gas expansion, and W33M2 as comprising moving stars and bow shocks. 
In this picture we model the complex triple tail internal structure of W33M2\textsubscript{a}, shown in Figure \ref{fig:w33_2+3_star_curves},  as tracing the intertwining of multiple stellar orbits. The simulations suggest that W33M2\textsubscript{a} includes 4 stars orbiting each other in a quaternary system (Star \#1-4) and one whose membership in the system is unclear (Star \#5), and that each of W33M3\textsubscript{a}, W33M3\textsubscript{b} and W33M2\textsubscript{b}  hold one star. 

\begin{figure*}[t]
	\gridline{\fig{W33M2+3_Orig_PVD_Dec.png}{0.54\textwidth}{(e)}
		\fig{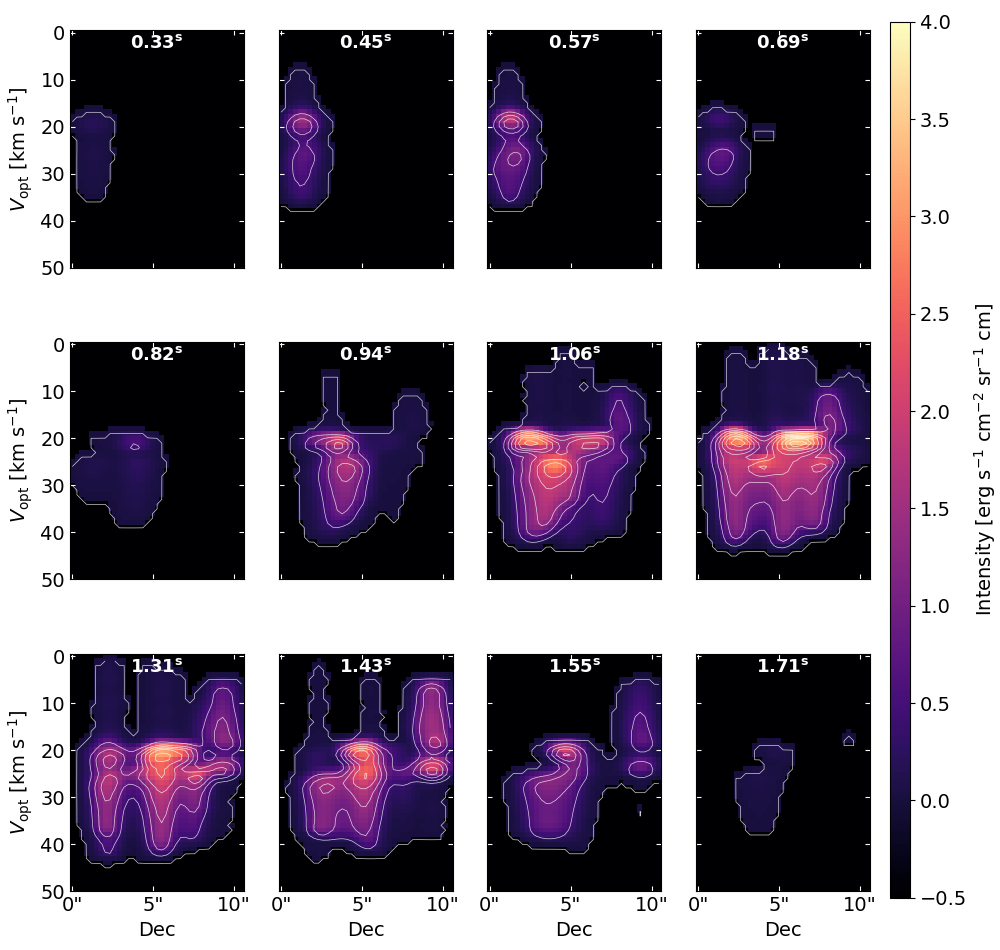}{0.54\textwidth}{(f)}}	
	\caption{Comparison of original data cube and simulation results for W33 Main 2+3: Channel maps of (a) W33 Main [NeII] data cube cropped with contour levels at [0.05, 0.1, 0.2, 0.5, 1.0, 1.5, 2, 2.5, 3] \flux~ and (b) Line profile emission simulation results with contour levels at [0.0, 0.5, 1, 1.5, 2, 2.5, 3.0, 3.3, 3.5, 3.7, 4.0] \flux. Position-Velocity Diagrams in RA from the data cube (c) and the simulations (d), Position-velocity Diagrams in Dec from the data cube (e) and the simulations (f). }
	\label{fig:w33_main_2_results}
\end{figure*}

In Figure \ref{fig:w33_main_2_results} we compare channel maps and PVDs of the original data on W33M2 and W33M3 to the simulations.  The channel maps show  reasonable agreement of the simulation results and the original data cube; both simulations and data show a ring with secondary clumps that peak at different velocities. As we saw for W33M1 the secondary features are more defined in the model than in the original data.  In both the data and the simulations,  PVDs in both dimensions display two bright 'arms', spatially distinct and extended in velocity, that join at a 'head' to form the bow shock, the shock appearing in the PVDs as a region of steep contours, and a spatially distinct tail extending to blue velocities  The simulated PVDs however are 'clumpier' than the original data, which is notably smooth. The 'clumps' in space and velocity show the presence of the multiple exciting stars assumed in this model; the relative smoothness of the original data may argue that fewer or only one star is present. 

\begin{figure*}[htp]
	\gridline{\fig{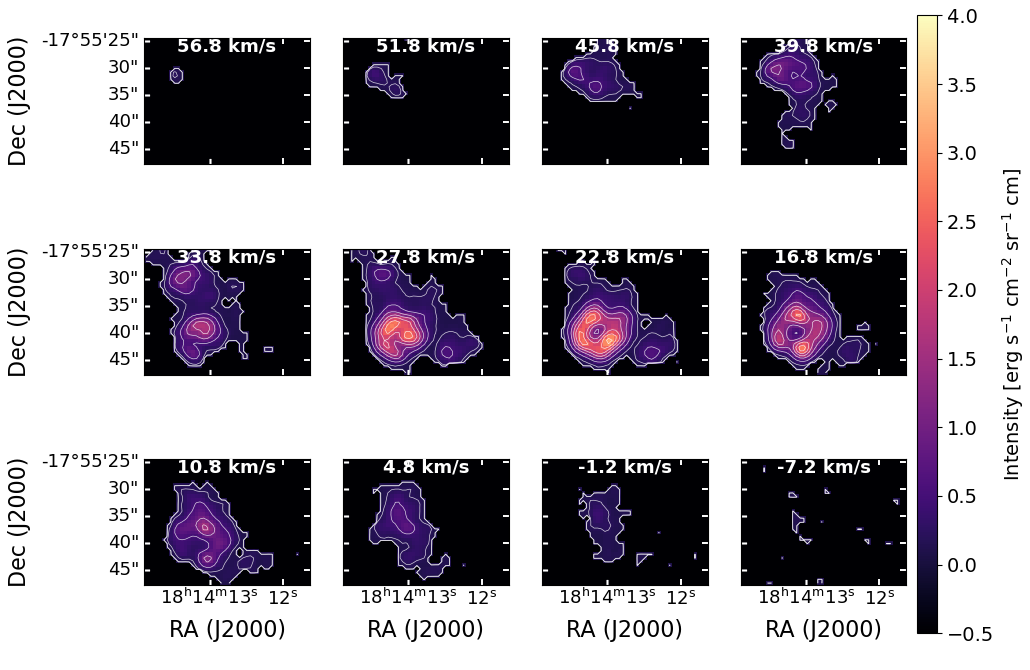}{0.54\textwidth}{(a)}
		\fig{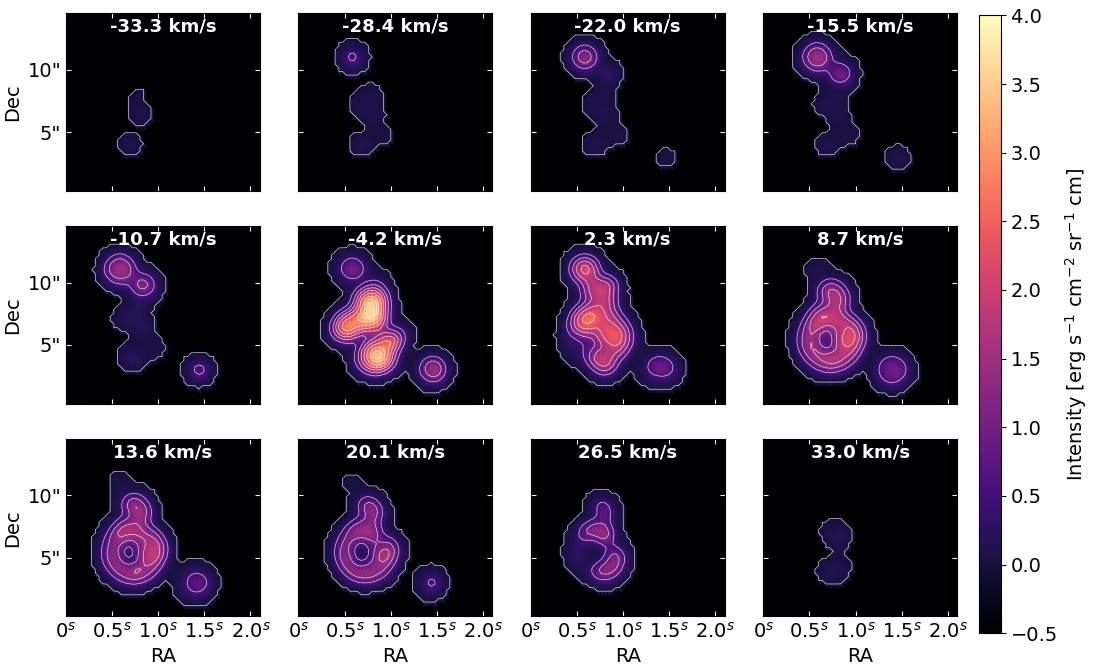}{0.54\textwidth}{(b)}}
\gridline{\fig{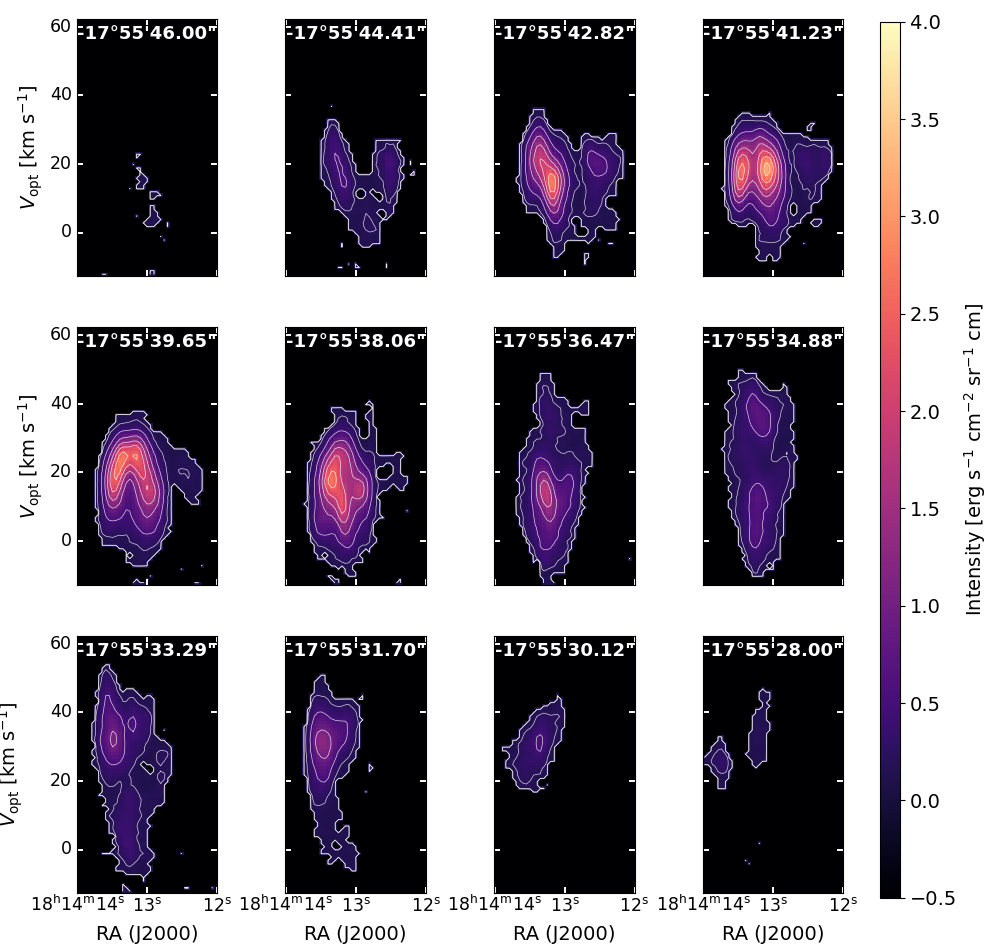}{0.54\textwidth}{(c)}
		\fig{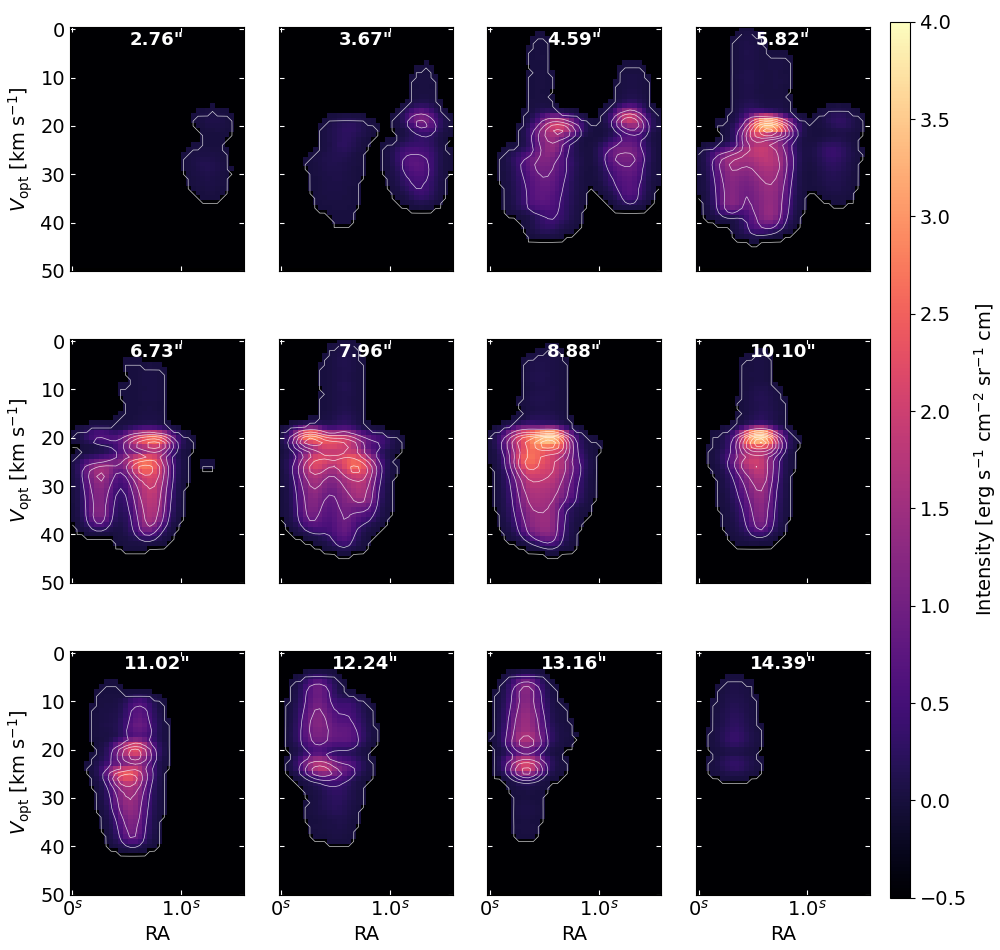}{0.54\textwidth}{(d)}}
		\vspace*{5cm}
\end{figure*}

\subsection{W33 Main 4} \label{sec:results_w33_main_4}
\begin{figure*}
\gridline{\fig{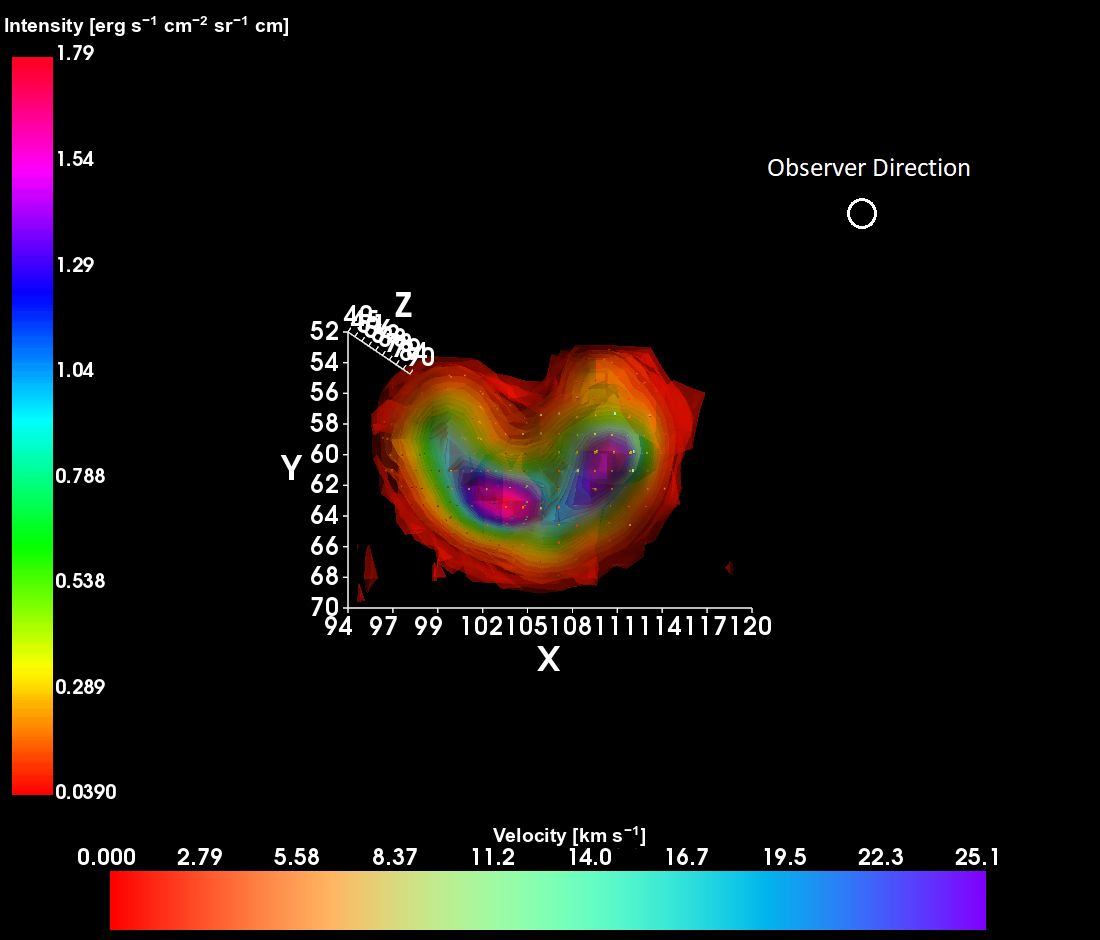}{0.37\textwidth}{(a)} 
       \fig{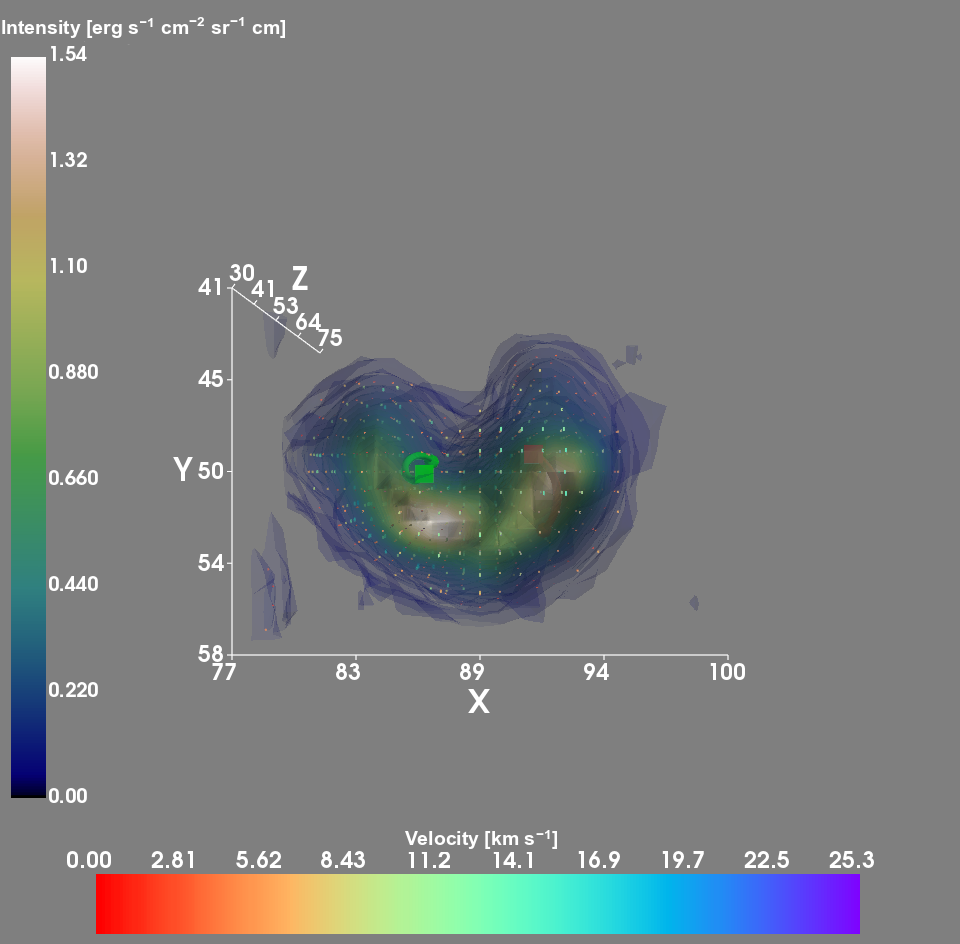}{0.37\textwidth}{(b)}}
\gridline{\fig{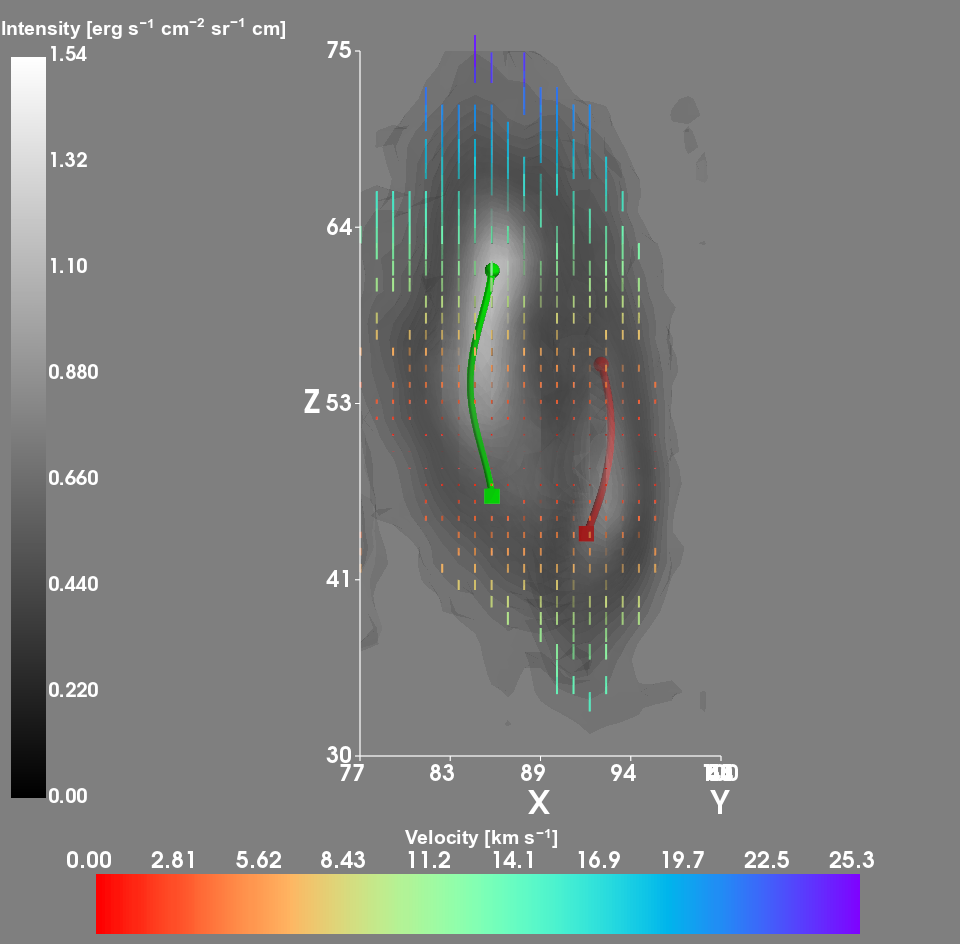}{0.37\textwidth}{(c)}
		\fig{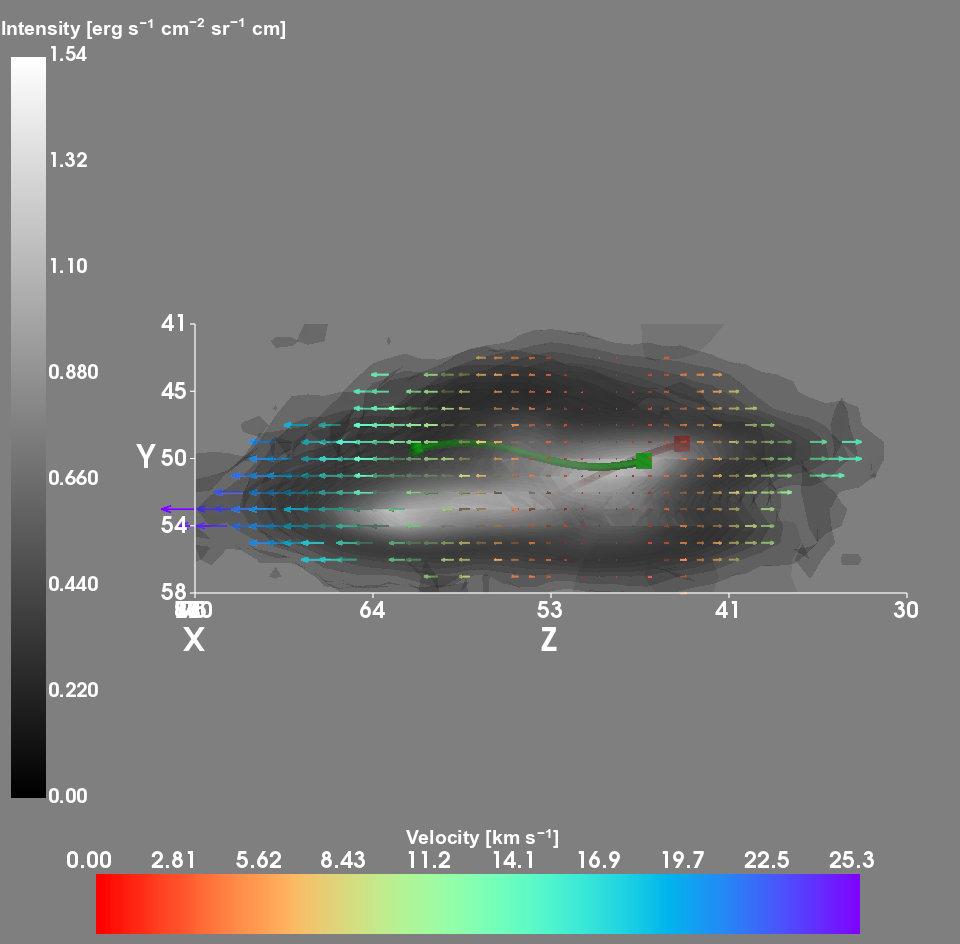}{0.37\textwidth}{(d)}}
       \gridline{\fig{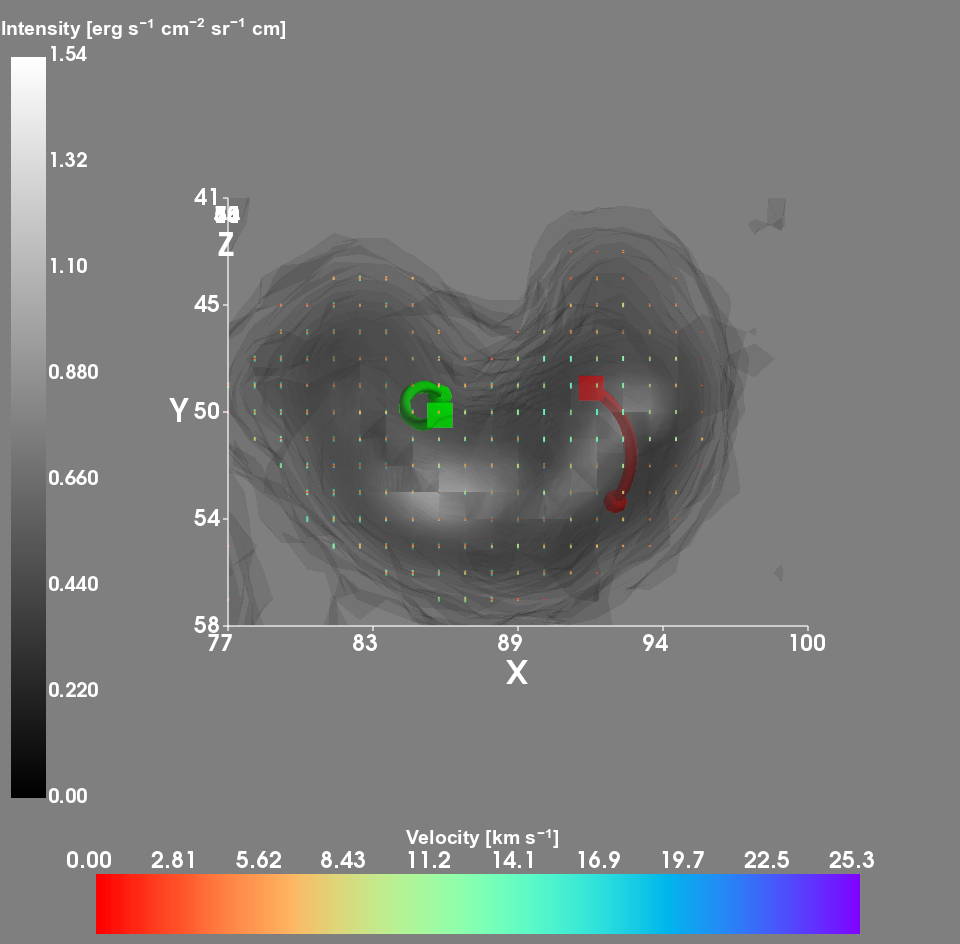}{0.37\textwidth}{(e)}
        \fig{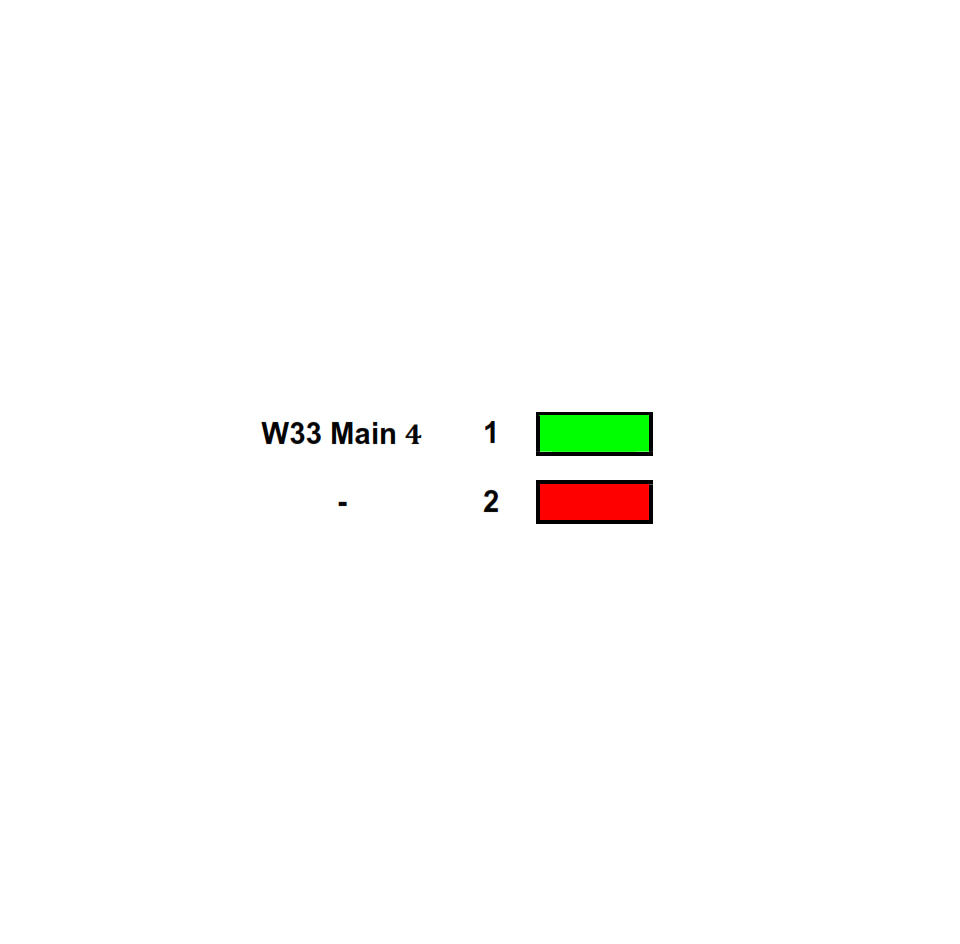}{0.37\textwidth}{}}
		\caption{3D View of the original [NeII] data cube cropped to the volume containing W33 Main 4 with the colored star movement curves (legend on bottom right) from 3 axes directions (Z- is the observer direction): (a) Dec, (b) RA, (c) Spectral/Kinematic. The X,Y,Z units are as follows: X axis is RA (1 X Unit = 0.66 arcsec), Y axis is Dec (1 Y Unit = 0.05s) and Z axis is spectral/kinematic axis (1 Z Unit = 1 \kms). The velocity axis is also denoted by the colored arrows where only the absolute velocity is considered. Each star curve shows the movement of a single star starting from the sphere shape and ending with the cube shape.}
		\label{fig:w33_4_star_curves}
\end{figure*}

W33M4 has the simplest structure of the UCHII regions: the radio measurements \citep{beck1998infrared} were consistent with a single embedded star.  The shape of W33M4 suggests a cometary UCHII region seen sideways but the kinematics do not agree:  the data shows two clumps offset in velocity by $\sim10$\kms~ with some interaction between them.   We found the the best fit to be a binary pair in an orbit at about $\sim 30^{\circ}$ relative to the observer.  The stellar trajectories are shown in Figure \ref{fig:w33_4_star_curves}(c,d,e). 

\begin{figure*}[htp]
	\gridline{\fig{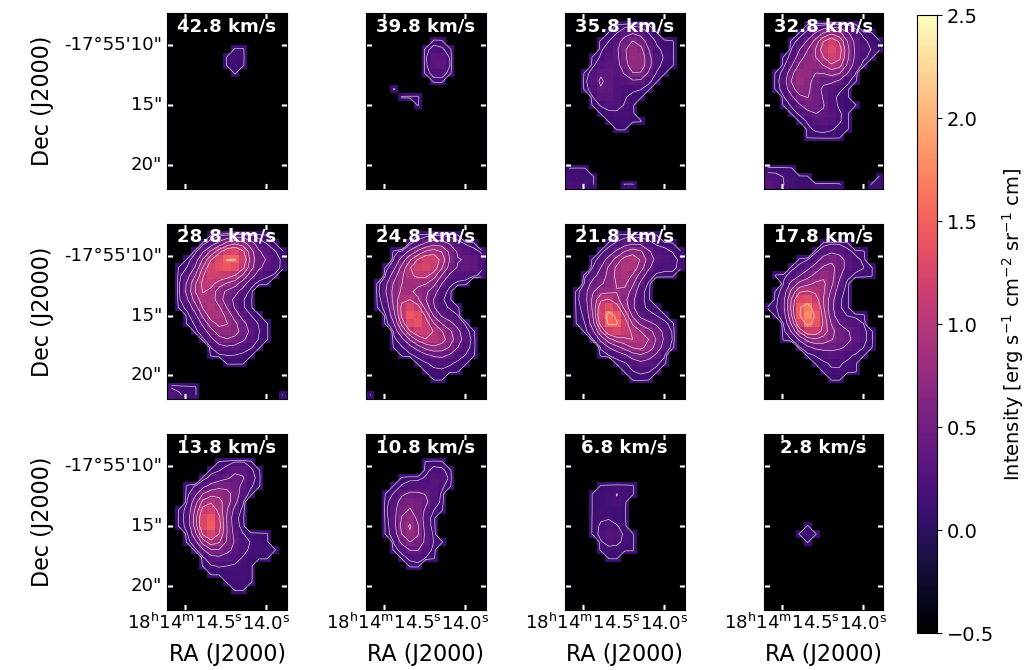}{0.54\textwidth}{(a)}
		\fig{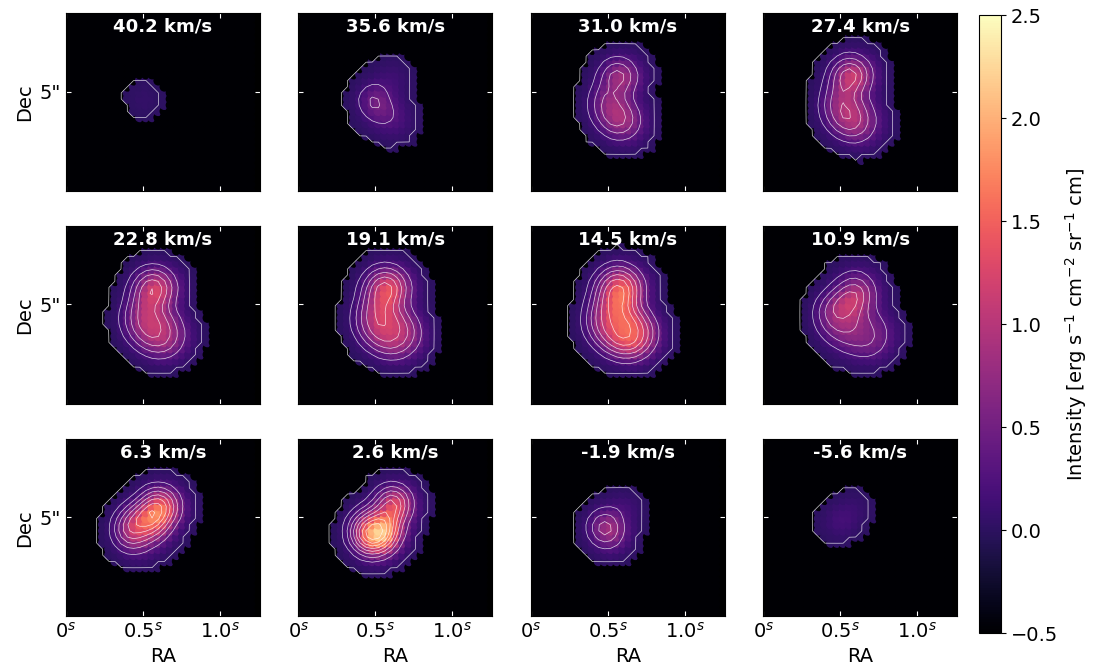}{0.54\textwidth}{(b)}}
\gridline{\fig{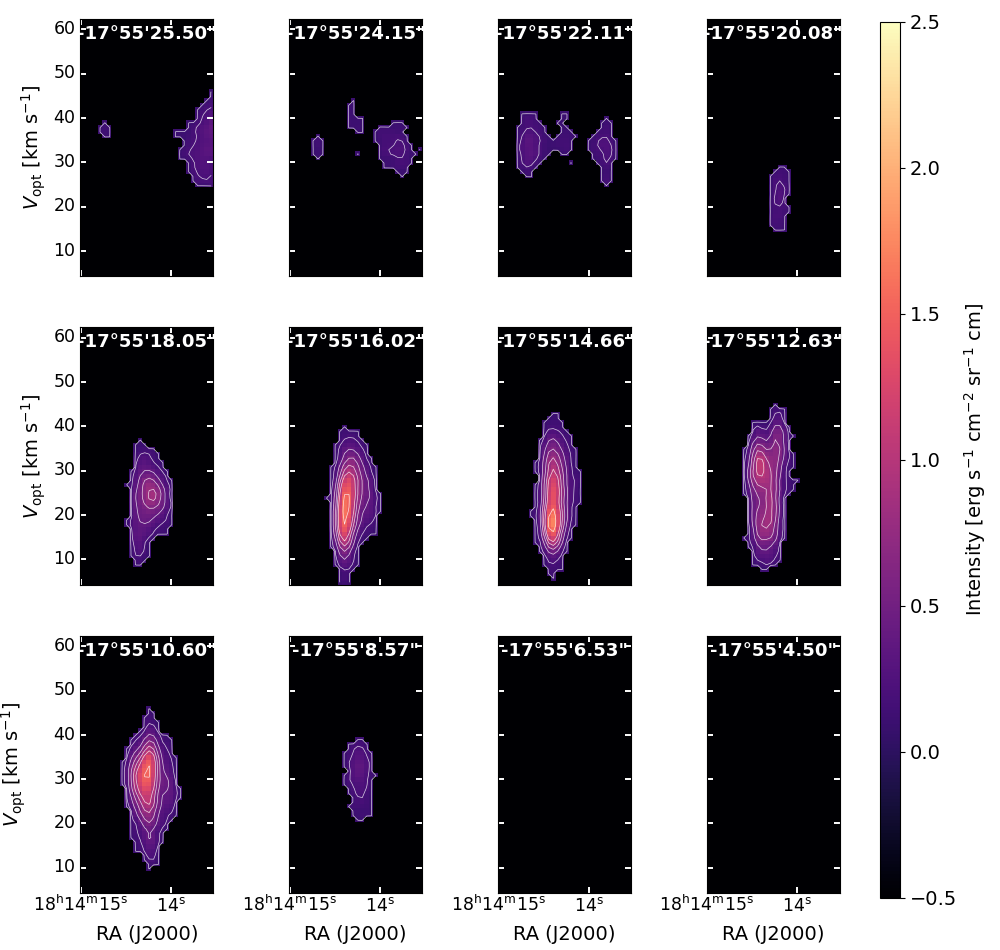}{0.54\textwidth}{(c)}
		\fig{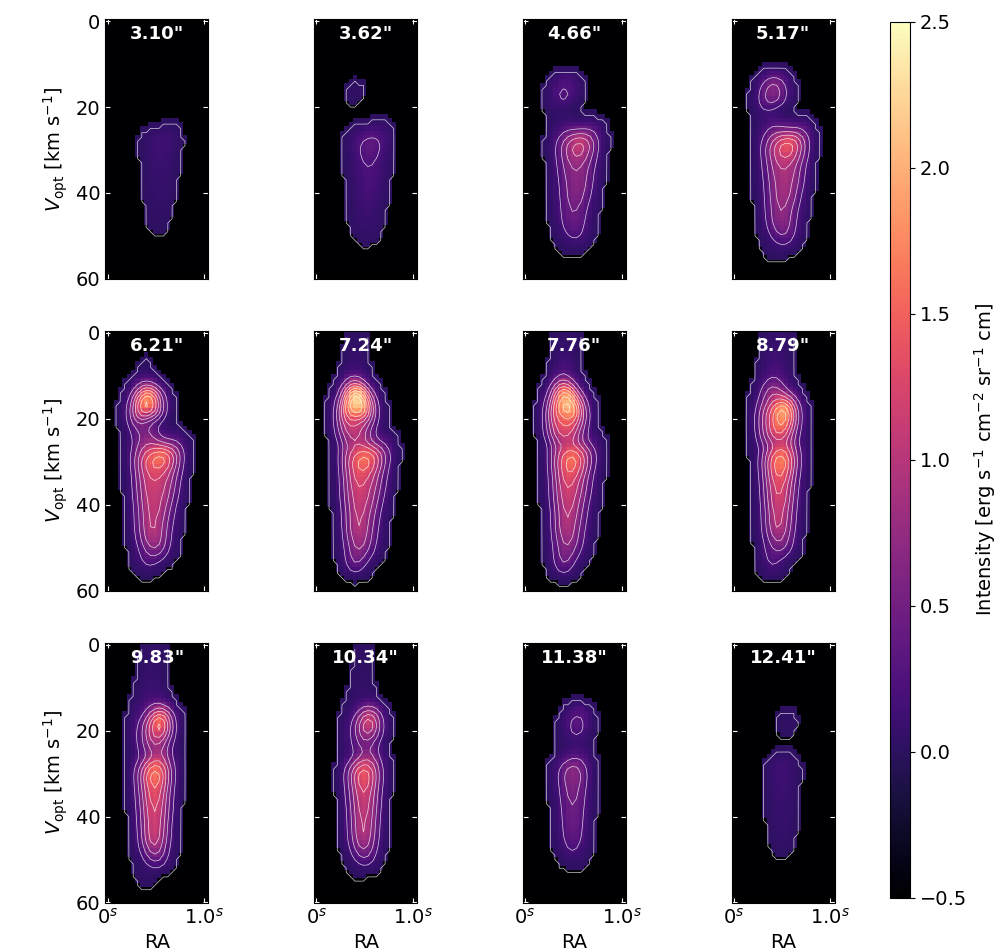}{0.54\textwidth}{(d)}}
			\vspace*{5cm}
\end{figure*}
\begin{figure*}[t]
	\gridline{\fig{W33M4_Orig_PVD_Dec.png}{0.54\textwidth}{(e)}
		\fig{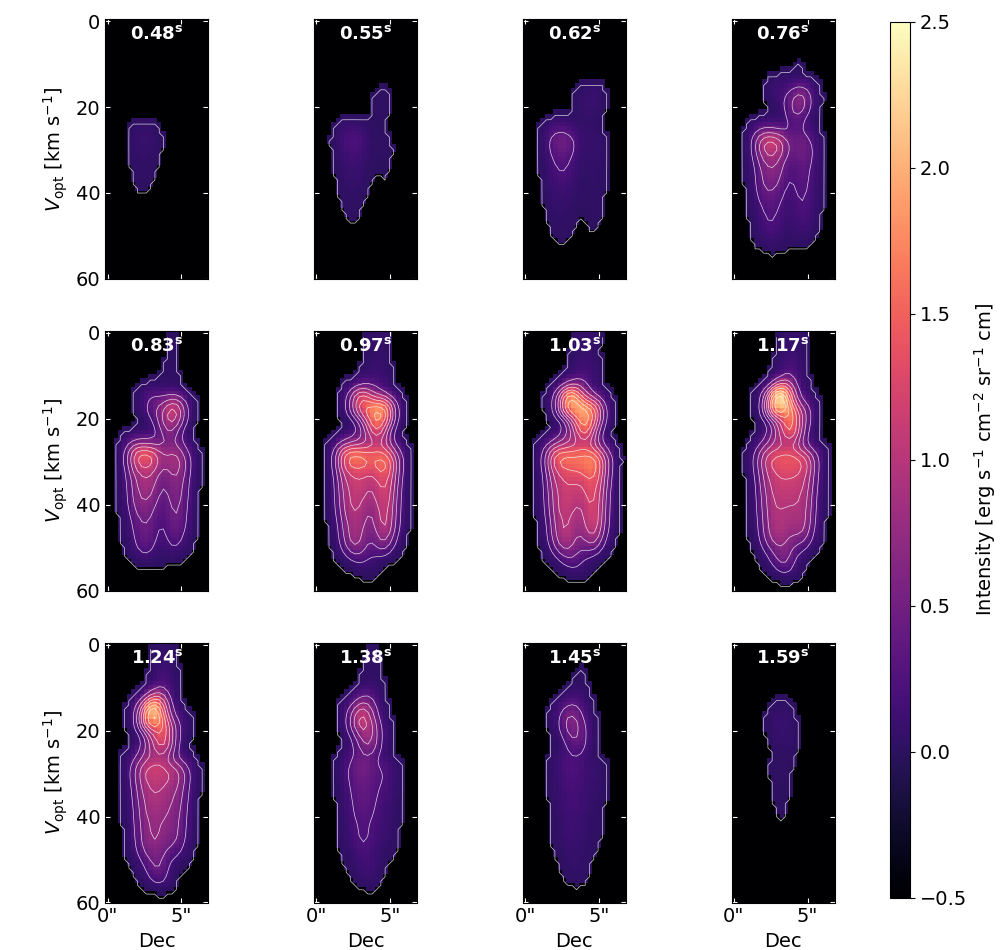}{0.54\textwidth}{(f)}}	
	\caption{Comparison of the original data cube and the simulation results for W33 Main 4. Channel maps of (a) the [NeII] data cube cropped with contour levels at [0.05, 0.1, 0.2, 0.5, 1.0, 1.5, 2, 2.5, 3] \flux and (B)  line profile emission simulation results with contour levels at [0.0, 0.5, 1, 1.5, 2, 2.5, 3.0, 3.3, 3.5, 3.7, 4.0] \flux. Position-Velocity Diagrams in  RA from the data cube (c) and the simulations (d). Position-Velocity Diagrams in Dec from the data cube (e) and the simulations (f).}
	\label{fig:w33_main_4_results}
\end{figure*}

In Figure \ref{fig:w33_main_4_results} we show the results of the hydrodynamic and radiative simulations in 3D for W33M4. The channel maps of the simulations agree fairly well with the data.   Both the original and simulated PVDs show two sources, offset in velocity by $\sim10$\kms~ and each extending $\sim20$\kms;
interaction and overlap between the two is more apparent in the simulations than in the original data.  

\section{The Simulated Stellar Population and Mass Function} \label{sec:results_population}

\begin{figure*}
	\gridline{\fig{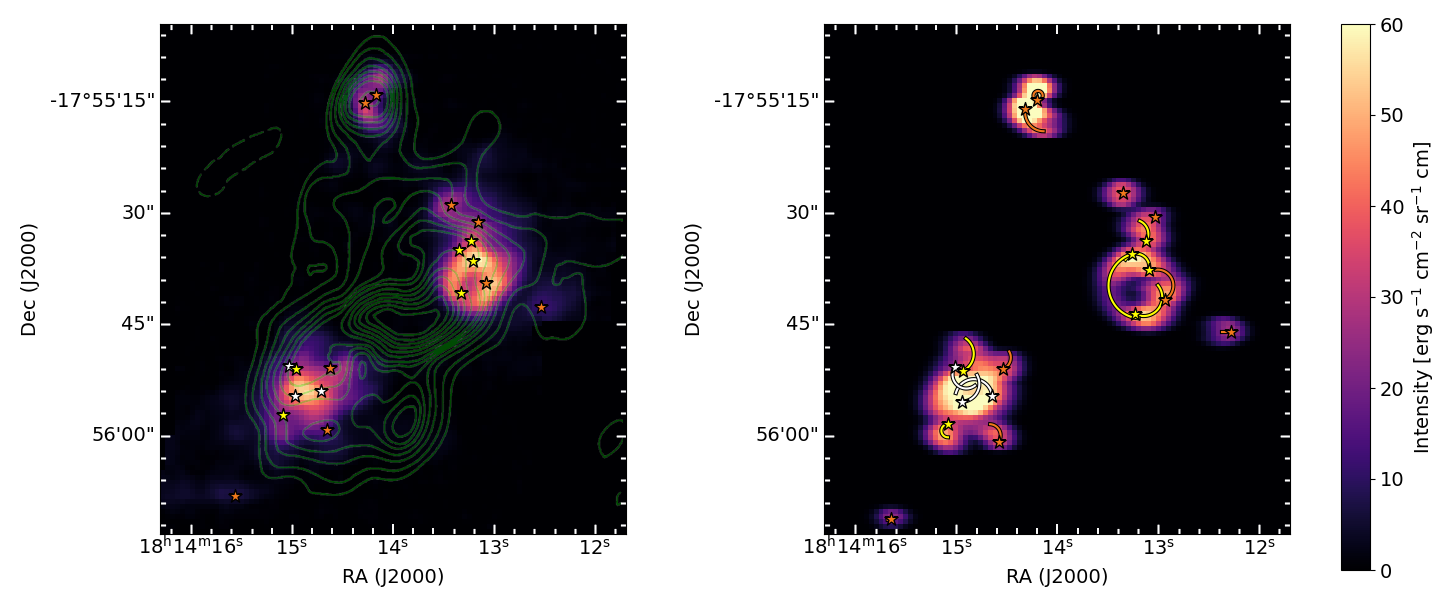}{1.0\textwidth}{}}
	\caption{Comparison of the 0\textsuperscript{th} moment map of the original W33 Main [NeII] data cube with the contours of the 6cm VLA radio map of W33 Main taken from \citet{haschick1983formation} (a) and the composite data cube (b) made up of all UCHII regions which we've simulated. On both maps the final position of each star in the simulations is denoted by colored star symbols, where their color indicates their mass ( B0.5 - white, B1V - yellow, B1.5V - orange). The simulated composite map also shows their approximate movement during the simulations.
		\label{fig:moment0_overlay}}
\end{figure*}

\begin{figure*}
	\gridline{\fig{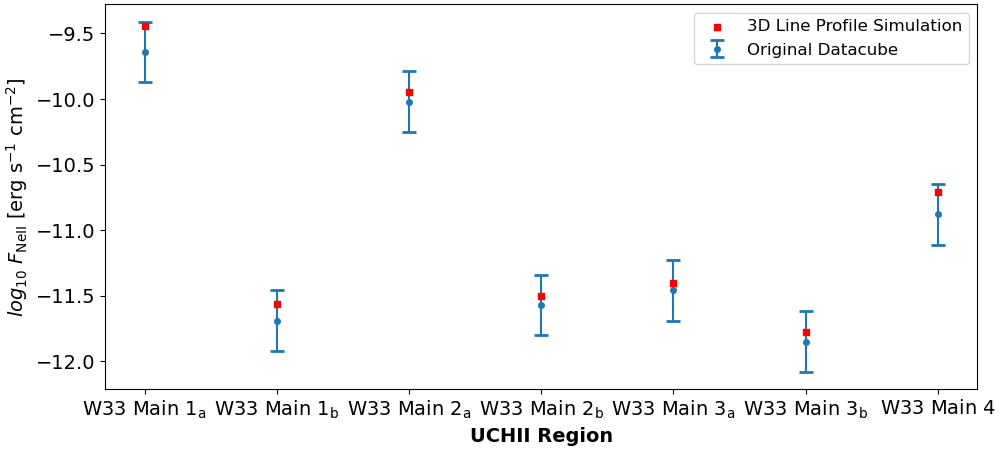}{0.7\textwidth}{}}
	\caption{Comparison of total [NeII] flux of each UCHII region in W33 Main between the original data cube (blue circles) and the 3D line profile simulation results (red squares).
		\label{fig:total_flux_comparison}}
\end{figure*}

  Our simulations propose multiple stars for each UCHII region in W33 Main. These stars, and their trajectories, are shown in in Figure \ref{fig:moment0_overlay}  overlaid on the 0\textsuperscript{th} moment maps of the original data cube and the 6cm radio map from \citet{haschick1983formation}; the models and the data agree spatially and kinematically. 
The deduced stellar masses are all of early B type. These are later than the stellar types quoted by \citet{beck1998infrared}, but there are substantial uncertainties and possible degeneracies in their derivation of stellar type from infrared line ratios.  Table \ref{tab:pluto_stars_parameters} shows that there are more stars in the lower mass than the high mass ranges, as predicted by all normal IMFs (initial mass function), but the number of stars is too small and the mass range covered too narrow for us to derive a meaningful exponent $\alpha$ for the local IMF. 
The total ionizing flux of the simulated UCHII regions is compared to the original data cube in Figure \ref{fig:total_flux_comparison};  most of the simulated UCHII regions slightly exceed the original in flux.  The simplest explanation of this discrepancy is that some of the 'stars' whose trajectories appear in the models are not true OB stars but dense clumps whose embedded stars, if any, are not ionizing.   

Are the high numbers of stars suggested by our simulations realistic?   In the empirical limit of high-mass star formation, proposed by \citet{kauffmann2010mass},  the number of stars in each UCHII region can be predicted from the total mass in a molecular clump:
\begin{equation}
	m(r) > 870 \text{M\textsubscript{\(\odot\)}} \left(\frac{r}{\text{pc}}\right)^{1.33}
\end{equation}
In Table \ref{tab:stars_num_comparison} we compare this prediction to the simulations. The predicted and simulated stellar populations agree with the exception of  W33M1\textsubscript{a},  another indication that some of the proposed stars in W33M1\textsubscript{a} are not true stars.
\begin{deluxetable*}{cccccccc}
	\tablenum{2}
	\tablecaption{Predicted and Simulated Stellar Populations \label{tab:stars_num_comparison}}
	\tablewidth{0pt}
	\tablehead{
		\colhead{UCHII Region} & \colhead{$R\textsubscript{region}$ [pc]} & \colhead{$M\textsuperscript{min}\textsubscript{region}$ [M\textsubscript{\(\odot\)}]} & \colhead{$N\textsubscript{predicted}$} & \colhead{$N\textsubscript{simulation}$}
	}
	\startdata
	W33 Main 1\textsubscript{a} & 0.13 & 60 & 4-5 & 7 \\
	W33 Main 1\textsubscript{b} & 0.04 & 12 & 1 & 1 \\
	W33 Main 2\textsubscript{a} & 0.11 & 49 & 3-4 & 5 \\
	W33 Main 2\textsubscript{b} & 0.05 & 16 & 1 & 1 \\
	W33 Main 3\textsubscript{a} & 0.05 & 18 & 1 & 1 \\
	W33 Main 3\textsubscript{b} & 0.03 & 9 & 1 & 1  \\
	W33 Main 4 & 0.07 & 25 & 2 & 2 \\
	\enddata
	\tablecomments{	Comparison of the predicted and simulated number of stars in each of W33 Main's UCHII regions, with $N\textsubscript{predicted}$ from the $M\textsubscript{region}$ of \citet{kauffmann2010mass} and $N\textsubscript{simulation}$ from our models.}
\end{deluxetable*}

\section{Conclusions and Discussion} \label{sec:conclusion}
 This paper reports on simulations of the UCHII regions of W33 Main which were motivated by the goal of explaining   high resolution data on the ionized gas.    We have obtained a [NeII] 12.8\um data cube of W33 Main with 1.4 arcsec spatial and $\sim5$\kms~ velocity resolution; this velocity resolution is a crucial advantage over \HI line measurements which are limited by thermal broadening to $\gtrapprox 21$\kms.~  Unlike most earlier work on UCHII regions which assumed single massive stars in each source, our simulations are based on the picture that each UCHII may contain several smaller ionizing stars, and that the bow shocks of these stars and their relative motions can create the complicated gas distributions and velocity features that are observed.  Our simulations include visualizations that permit us to fine-tune the stellar trajectories to  match the spatial and kinematic effects observed.   We found that:
\begin{itemize}
 \item  The structures and kinematics observed for these UCHII regions cannot,  with the one exception of W33M2\textsubscript{a}, be satisfactorily modelled by a single embedded star either moving or stationary.  
\item  Each UCHII region can be simulated by a range  of lower mass OB stars, moving relative to each other;  the stars,  the winds and their interactions interweave to create the complex kinematic and spatial structures seen in the data.
\item The total stellar population suggested by the simulations is dominated by the lower-mass stars.
\item  The total ionization of the suggested stellar population slightly exceeds the observed [NeII] result, suggesting that some of the proposed 'stars' are density clumps or low mass stars, rather than ionizing OB stars.  The worse discrepancy is in the complex source W33M1 and we think it possible that some of the 'stars' in the outer regions of that source are in fact non-ionizing.  It may be that the central stars and gas are interacting to produce gas arms rather than the usual cometary shape on which the simulations are based (\citep{wang2012multiple} or \citep{homan2015simplified}).
\end{itemize}

We have shown that models with multiple moving stars produce complex spatial and kinematic features, and have created such models consistent with the observations of W33 Main. But as in any simulation study, 'consistent with the observations' does not necessarily mean that the model is correct.   A given spatial and velocity feature may have been created by any of several mechanisms.  W33M2\textsubscript{a} is a cautionary example.   We have presented here models in which the many density features in W33M2\textsubscript{a} (Figure \ref{fig:w33_2+3_star_curves}) reflect structures in the ambient cloud or the wind.  We interpreted these complicated internal structures as the 'curved tails'  created by multiple stars in orbit, and the simulations show that this morphology can be created by a system of 4 stars.   However the channel maps and PVDs  of W33M2\textsubscript{a} also resemble the cometary models of \citet{Zhu_2015} in which there is a single stationary ionizing star and the two 'arms' of the PVD show gas flowing out of the UCHII region along the walls of a cavity.  In those cometary models the ionized gas is predicted to be blue-shifted with respect to the embedding cloud; note that  W33M2\textsubscript{a} is $\sim10$\kms~ blue of the the molecular cloud velocity and of the other UCHII regions.  Further observations with high resolution could determine which model, or combination of models, is most appropriate for this source.

Our results have suggested that the appearance and kinematics of  UCHII regions may be created by interactions of multiple late-type OB stars. This picture can readily be checked with sub-arcsecond resolution radio mapping. If confirmed it will give a more accurate stellar population for the star formation region and a clue to the initial mass function.  Another intriguing result of these simulations is that they point to the relative motions of the stars in each UCHII region. If confirmed, this model can show how the stellar proto-cluster W33 Main will develop, the trajectories of the stars and how the stars will disperse. 
 
Finally, these results show the value not only of the simulations  but of the data cube: the results depend on the very high spectral resolution possible with infrared  lines of metal ions.  The [NeII] data cube in W33 Main shows that the ionized gas has full velocity extent of only $\sim40$\kms~ and holds important kinematic features offset from each other by $\lesssim10$\kms;  it could not be studied properly under the $\sim20$\kms~ thermal broadening inevitable for an \HI recombination line .   This demonstrates the value of the metal ionic lines and of high spectral resolution in UCHII regions. 
\facilities{JHL was a Visiting Astronomer at the Infrared Telescope Facility, which is operated by the University of Hawaii under contract 80HQTR19D0030 with the National Aeronautics and Space Administration.}
\section{Appendix}
The 3D visualization program is available at \url{https://github.com/danbeilis/fits_stellar_movement_viz}. 
It offers an alternative method of analyzing the structure of data cubes by showing the structure in 3D. It makes it easy to examine different regions by limiting the range of the data shown;  either by cropping the axes or selecting the intensity range.  The program also makes it possible to add 2D slices along each of the three axes for a clearer view of the density structure.  The program has the option of using  arrows colored according to velocity to display kinematics without using the velocity axis. The 3D display changes instantly on changing the input parameters, which makes it simple to trace the trajectories of the stars and how they impact the UCHII regions.  
\section{Data Availability}
The [NeII] data cubes are available in FITS format at \url{https://github.com/danbeilis/data/tree/master/W33}. 

\bibliography{W33}{}
\bibliographystyle{aasjournal}



\end{document}